\documentclass[useAMS,usenatbib]{mn2e}

\usepackage{graphicx}
\newcommand{\arcm}{\hbox{$^\prime$}}
\newcommand{\arcs}{\mbox{\arcm\hskip -0.1em\arcm}}

\title[X-ray observations of three young, early-type galaxies]{X-ray observations of three young, early-type galaxies}
\author[A. E. Sansom et al.]{A. E. Sansom$^{1}$\thanks{E-mail:
    AESansom@uclan.ac.uk (AES)}, E. O'Sullivan$^{2}$, Duncan
  A. Forbes$^{3}$, R. N. Proctor$^{3}$, D. S. Davis$^{4}$ \\
$^{1}$Centre for Astrophysics, University of Central Lancashire, Preston, 
Lancashire, PR1 2HE, UK\\
$^{2}$Harvard-Smithsonian Center for Astrophysics, 60 Garden Street, Cambridge
MA 02138, USA \\
$^{3}$Centre for Astrophysics \& Supercomputing, Swinburne University, Hawthorn, VIC 3122, Australia \\
$^{4}$Joint Center for Astrophysics, Department of Physics, University of Maryland, Baltimore County, MD21250; \\
Laboratory for High Energy Astrophysics, NASA, GSFC, Code 662, Greenbelt, MD 20771, USA}

\begin{document}

\date{Accepted 2--- December 15. Received 2--- December 14; in original form 2--- October 11}

\pagerange{\pageref{firstpage}--\pageref{lastpage}} \pubyear{2006}

\maketitle

\label{firstpage}

\begin{abstract}
Massive halos of hot plasma exist around some, but not all elliptical galaxies.
There is evidence that this is related to the age of the galaxy.
In this paper new X-ray observations are presented for three early-type 
galaxies that show evidence of youth, in order to investigate their X-ray 
components and properties. NGC 5363 and NGC 2865 were found to have X-ray 
emission dominated by purely discrete stellar sources. Limits are set on the 
mass distribution in one of the galaxies observed with XMM-Newton, NGC 4382, 
which contains significant hot gas. We detect the X-ray emission in NGC 4382 
out to 4r$_e$. The mass-to-light ratio is consistent with a stellar origin 
in the inner regions but rises steadily to values indicative of some dark 
matter by 4r$_e$. These results are set in context with other data drawn 
from the literature, for galaxies with ages estimated from dynamical or 
spectroscopic indicators. Ages obtained from optical spectroscopy represent 
central luminosity weighted stellar ages. We examine the
X-ray evolution with age, normalised by B and K band luminosities. 
Low values of Log(L$_X$/L$_B$) and Log(L$_X$/L$_K$) are found for all 
galaxies with ages between 1 and 4 Gyrs. Luminous X-ray emission only 
appears in older galaxies. This suggests that the interstellar medium 
is removed and then it takes several gigayears for hot gas halos to build 
up, following a merger. A possible mechanism for gas expulsion might be 
associated with feedback from an active nucleus triggered during a merger.

\end{abstract}

\begin{keywords}
X-rays: galaxies --galaxies: evolution -- galaxies: ISM -- 
galaxies: individual: NGC 4382, NGC 5363, NGC 2865
\end{keywords}

\section{Introduction}

The existence of hot gas halos around many elliptical galaxies presents 
something of an enigma. How were they created and why do they not exist 
in all cases? Supernovae are supposed to provide the energy to heat up 
an existing gas reservoir, which could originate from stellar mass loss 
or from externally accreted gas. This heating process would require a large 
amount of energy to heat the quantities of gas, detected in X-rays,  
to the required temperatures (few $\times$ 10$^6$ K). Once heated, the gas 
may form a hot halo around the 
galaxy, or escape the potential well of the galaxy in a wind, or take part 
in cooling flows, particularly rapid near the denser central regions of 
the galaxy (Pellegrini \& Ciotti 1998; Fabian et al. 2002; Daisuke, 
Yun \& Mihos 2004). 
Another possibility for the gas origin is from the cold interstellar 
medium (ISM) in spiral galaxies if the hot gas halos formed when they 
merged. However, previous work showed that there is an unexpected dearth 
of gas, either cold or hot, in young, early-type galaxies (E and S0 galaxies). 
Results from the Einstein and ROSAT satellites indicated that dynamically 
young, early-type
galaxies are X-ray poor compared to other early-type galaxies (Fabbiano 
\& Schweizer 1995; Sansom, Hibbard \& Schweizer 2000; O'Sullivan, Forbes 
\& Ponman 2001a, hereafter OFP01a). Hibbard \& Sansom (2003) studied 5 
of these dynamically young, early-type galaxies in neutral hydrogen (HI), 
with the VLA and 
found very low upper limits to the mass of cold gas present in these 
systems ($<2\times 10^7 M_\odot$). Chang et al. (2001) found a similar result 
for E+A galaxies, thought to be recent merger products, with only one out of
5 galaxies detected in HI. Georgakakis et al. (2001) showed that the HI 
content of young merger remnants decreases in the 
first 1 to 2 Gyrs following a merger, after which there is little detectable 
change in HI content towards evolved ellipticals. More recently 
Xilouris et al. (2004a) studied the dust content along a sequence of merging 
systems with dynamically estimated ages. They found that the warm-to-cold 
dust mass ratio increased along this sequence indicative of changing star 
formation and dust content in ongoing mergers. 

OFP01a looked at X-ray emission in 
early-type galaxies versus their luminosity weighted age estimated 
from optical spectroscopy (Terlevich \& Forbes 2002, hereafter TF02). 
From optical spectroscopy the luminosity weighted age is likely to 
indicate the age of the last major gaseous merger/accretion, since the 
luminosity weighting is dominated by the latest star formation. 
OFP01a showed that ellipticals with younger luminosity weighted ages were 
generally weaker X-ray sources, when normalised by their optical luminosity. 
Note that optical luminosity fades only slowly for stellar populations 
older than about a Gyr, as we illustrate later in this paper. 

These previous studies made use of simplistic characterisations of the
overall X-ray properties of early-type galaxies, since the signal-to-noise
or spectral sampling of the X-ray observations was generally low. With the
advent of more sensitive X-ray missions, such as ASCA, XMM-Newton and
Chandra, the increased signal-to-noise and broader spectral range allows us
to investigate the X-ray properties of early-type galaxies in more detail.
ASCA observations revealed that elliptical galaxies generally require two
spectral components to describe their X-ray emission (Matsushita et al.
1994; White, Sarazin \& Kulkarni 2002). These originate from diffuse gas
(described by a soft thermal component at kT$\sim$ 0.2 to 1 keV) and from a
population of low-mass X-ray binary (LMXB) stars (with higher energy
emission that can be described by a power-law with flux $\propto$
frequency$^{-1.8}$).  In this paper we focus on three early-type galaxies
(NGC 4382, NGC 5363 and NGC 2865) which are thought to be young. The aim is
to measure the levels of different contributions to the X-ray emission.  In
particular, we focus on determining the contributions from hot gas. 

In this paper we employ galaxy 'ages'. Briefly, such ages are derived from 
a longslit, optical spectrum of the galaxy centre. The Lick system 
(e.g. Trager 2004 and references therein) defines 25 absorption lines 
that can be measured from the spectrum and corrected onto the same basis as
the original work (e.g. Worthey 1994). Comparison of the resulting indices 
to single age, single metallicity stellar populations can break the 
age-metallicity degeneracy to give independent, luminosity-weighted 
age and metallicity estimates. In our case we use the multi-line $\chi^2$ 
fitting method of Proctor \& Sansom (2002).

It should be recognised that the resulting ages are applicable only to 
the central regions, typically 1/8 to 1/2 of the galaxy effective radius 
(r$_e$). In general galaxy centres will contain a complex mix of stellar 
population ages. Any young stellar population present will be brighter than
an old population. Thus the derived luminosity-weighted age for the galaxy 
centre will infact be an upper limit to the mean age of the central stars.
The youngest stars were formed in a starburst event that may have been 
triggered by gas accretion onto the galaxy centre by an interaction or 
merger. Forbes, Ponman \& Brown (1998) showed that the spectroscopic age 
was similar to the time since a merger for a small sample of morphologically 
disturbed galaxies. Thus such ages provide an indication of the time since 
the last interaction or merger, but not necessarily about when the bulk of 
the galaxy's stars formed. This will depend on the (unknown) fraction of 
mass involved in the young stellar population.

This paper is set out in the following way. Section 2 describes the targets 
and their observations. Sections 3 presents their X-ray properties derived 
from spectral fitting of XMM-Newton and Chandra data. Sections 4 describes
analysis of the X-ray surface brightness profile and limits on the mass
distribution in NGC 4382, from combined XMM-Newton and archival Chandra
data. Section 5 sets these results into broader context with X-ray data
from other early-type galaxies that also have age estimates. Conclusions 
are given in Section 6.

Throughout this paper, we normalise optical $B$-band luminosities to the
$B$-band luminosity of the sun, L$_{B\odot}$=5.2$\times$10$^{32}$ ergs 
s$^{-1}$, and assume H$_0$=75~km~s$^{-1}$~Mpc$^{-1}$. Abundances are
measured relative to the ratios of Anders \& Grevesse (1989). While these
have now been superseded by more recent measurements, their use provides
continuity with previous studies.

\section[]{Targets and observations}

We aimed for a sample of early-type galaxies with evidence of young ages in
order to study their X-ray emission components in detail. Targets were
selected from the X-ray catalogue of early-type galaxies of O'Sullivan,
Forbes \& Ponman (2001b, hereafter OFP01b). The generally faint nature of
the X-ray fluxes from early-type galaxies means that sensitive
instrumentation is required to measure the various contributions to the
X-ray flux, from spectral fitting. A broad spectral range and good spectral
resolution are also required. The most sensitive X-ray observatory
currently available is XMM-Newton. Therefore we selected four nearby cases
thought to be young, early-type galaxies and were awarded time for two of
them on XMM-Newton, NGC~4382 and NGC~5363. We also made use of an archival
Chandra observation of NGC~4382 and an additional young galaxy, NGC 2865,
observed with Chandra is also reported in this paper. These three galaxies
are described below and their optical properties are summarised in Table~1.

\subsection{NGC 4382}
The X-ray luminosity of NGC 4382 was known from previous observations. 
EINSTEIN observations showed it to be a moderate luminosity X-ray source 
with (Log(L$_X$)=40.33 erg s$^{-1}$ for $H_0$ = 75 km s$^{-1}$ Mpc$^{-1}$, 
Fabbiano, Kim \& Trinchieri 1992). It was then observed as an X-ray 
faint, early-type galaxy, with ROSAT (Fabbiano, Kim \& Trinchieri 1994).
More recently it was observed with Chandra by Sivakoff, Sarazin \& Irwin 
(2003), who resolved 58 point sources within the galaxy, attributed mostly 
to LMXBs. They also detected some diffuse gas at kT$\sim 0.3$ 
and uncertain abundance. NGC 4382 is a lenticular galaxy that follows a 
de Vaucouleurs (r$^{1/4}$ law) optical surface brightness profile (Baggett, 
Baggett \& Anderson 1998). No neutral hydrogen 
gas is detected in this galaxy (Hibbard \& Sansom 2003). It is interacting 
with NGC 4394, both in the Virgo cluster. It has a large quantity
of morphological fine structure, which points towards a dynamically young age 
(Schweizer \& Seitzer 1992). The luminosity weighted age of its 
stellar population was also 
estimated to be young (1.6$\pm$ 0.3 Gyr) from optical spectral absorption 
lines, as listed in the Age Catalogue of Terlevich \& Forbes (TF02), see 
http://astronomy.swin.edu.au/dforbes. Its deviation from the 
fundamental plane also suggests it is very young (Forbes, Ponman \& Brown 
1998).

\subsection{NGC 5363} 
Results from EINSTEIN observations of NGC 5363 reveal a moderate X-ray 
luminosity of Log(L$_X$)=40.14 erg s$^{-1}$ (OFP01b). This galaxy has been 
classified in various ways, including irregular and peculiar. However 
morphological classification based on recent, mid-infrared maps gives an 
E/S0pec class (Pahre et al. 2004) and the galaxy follows an r$^{1/4}$ profile 
(Xilouris et al 2004b). NGC 5363 is a non-interacting pair with NGC 5364, 
which is 14.5 arcmins away. NGC 5363 was thought to be a young system, from 
its strong H$\beta$ absorption line. From optical spectroscopy Denicol\'{o} 
et al. (2005) estimate the age in the central regions to be 3.8 
$^{+2.1}_{-3.5}$ Gyr. They used only four spectral line-strengths to 
estimate the age (H$\beta$ versus the composite index [MgFe]). Later in this 
paper we attempt a new age estimate using more data (see Section 5.1).
NGC 5363 has a dust lane along its minor axis, 
which shows up in mid-infrared observations (Xilouris et al. 2004b). From 
observations with the Infrared Space Observatory (ISO) it contains a dust 
mass of $\sim 2\times 10^6$ M$_{\odot}$ (Temi et al. 2004). If the gas-to-dust 
mass ratio is similar to that in our Galaxy, this implies a total gas mass 
of at least $\sim 2\times 10^8$ M$_{\odot}$ in the ISM. The gas-to-dust mass 
ratio in early-type galaxies may be more than this ($\sim$3000, with large 
scatter, from the ISM catalogue of Bettoni, Galletta \& Garcia-Burillo 2003). 
With this gas-to-dust ratio the total gas mass in NGC 5363 could be 
$\sim 6\times 10^9$ M$_{\odot}$. We discuss qualitatively whether 
this is detected in the X-ray observations in Section 3.1 and Section 6.

\subsection{NGC 2865}
There was an upper limit on the X-ray flux from NGC 2865, of $<1.9\times
10^{-13}$ erg s$^{-1}$ cm$^{-2}$, from ROSAT all sky survey data (OFP01b), 
implying a luminosity limit of Log(L$_X$)$<40.48$ erg s$^{-1}$.
NGC 2865 is an isolated elliptical galaxy that follows an r$^{1/4}$ law 
in surface brightness, plus is surrounded by shells (e.g. Reda et al. 2004). 
It also has a kinematically distinct core (Hau et al. 1999). It is
quite blue, has strong H$\beta$ absorption of 3.12\AA \, and an estimated
age since the last major burst of star formation of between 0.4 and 1.7 Gyr 
(Hau et al. 1999), from fitting optical spectroscopy with star formation 
histories.  Despite its apparent youth NGC 2865 is classified as an 
elliptical (T-type -5) in de Vaucouleurs et al. (1991, hereafter RC3).

\bigskip
\begin{table*}
\centering
\begin{minipage}{150mm}
\caption{\bf Optical parameters for three young, early-type galaxies.
The columns give the galaxy name, morphological type and T-type, distance and 
redshift. The isophotal diameter (D25) is given next, then the half-light 
radius (r$_e$), total apparent B magnitude and estimated age of each galaxy.
Sources of information are indicated below column headings, where: 
NED = NASA Extragalactic database, 
RC3 = de Vaucouleurs et al. (1991),
PS96 = Prugniel \& Simien (1996).}
 
\begin{tabular}{lcccccccc}
 & & & & & & & & \\
{\bf Galaxy} &{\bf Type} &{\bf T-type} &{\bf Dist} &{\bf z} &{\bf D25} &{\bf r$_e$} &{\bf B$_T$} &{\bf Age} \\
 & & & {\bf (Mpc)} & &{\bf (``)} &{\bf (``)} &{\bf (mag.)} &{\bf (Gyr)} \\
 &NED &RC3 &PS96 &NED &RC3 &RC3 &RC3 &(See text) \\
 & & & & & & & \\
NGC 4382  &S0pec    &-1.0   &15.9     &0.00243 &425 &54.6 &10.00  &1.6  \\
NGC 5363  &E/S0pec$^a$ &-3.5$^a$  &15.8$^b$ &0.00380 &244 &36.1  &11.05  &3.8  \\
NGC 2865  &E3-4     &-5.0   &36.5     &0.00876 &147 &12.5  &12.57  &1.0 \\
\multicolumn{8}{l}{$^a$Pahre et al. (2004)} \\
\multicolumn{8}{l}{$^b$ Tully (1988)} \\
 & & & & & & & \\
\end{tabular}\\
\end{minipage}
\end{table*}

\begin{figure}
\includegraphics[angle=0,width=90mm,trim=0 0 0 0]{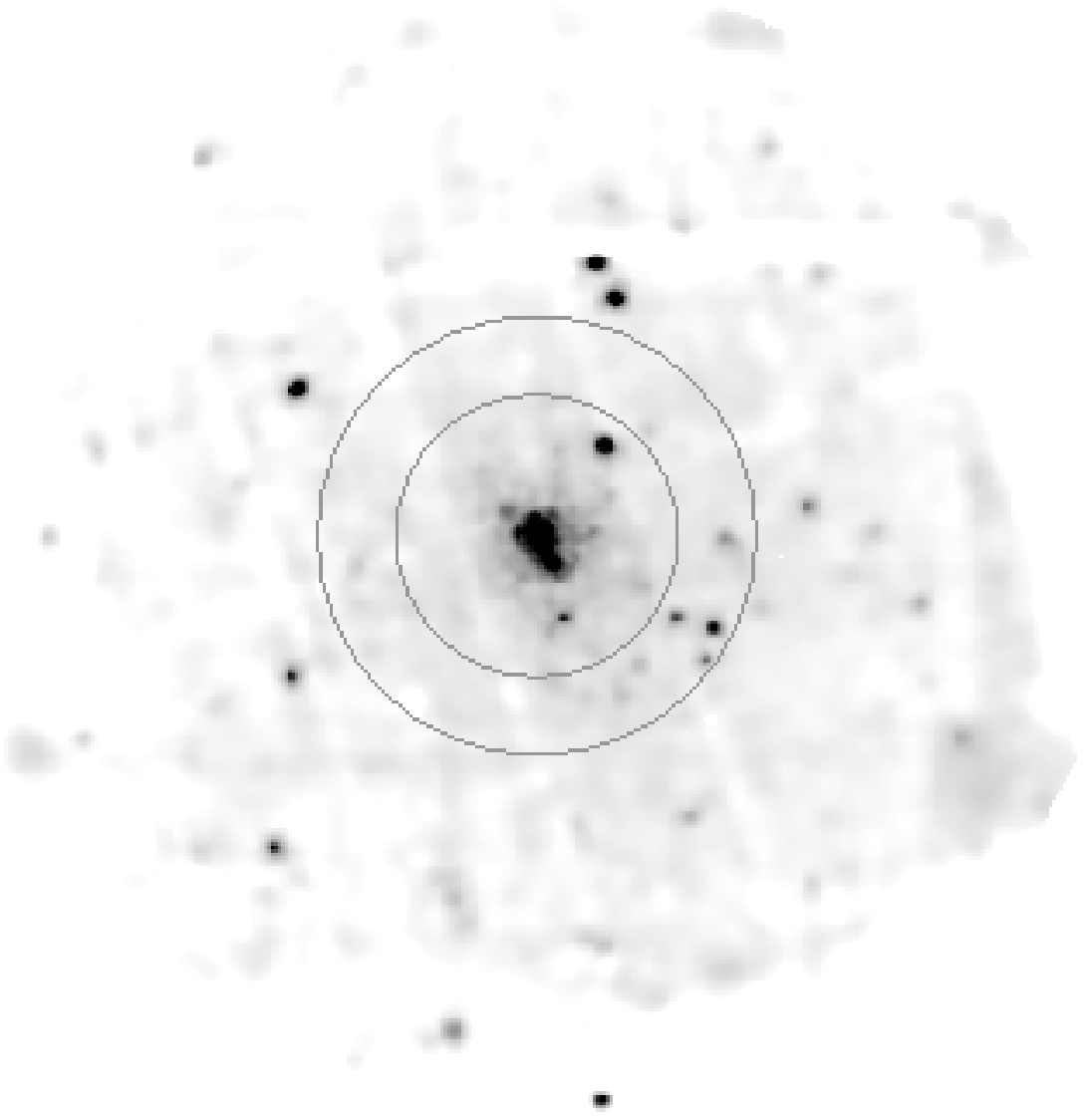}
\includegraphics[angle=0,width=90mm,trim=0 0 0 0]{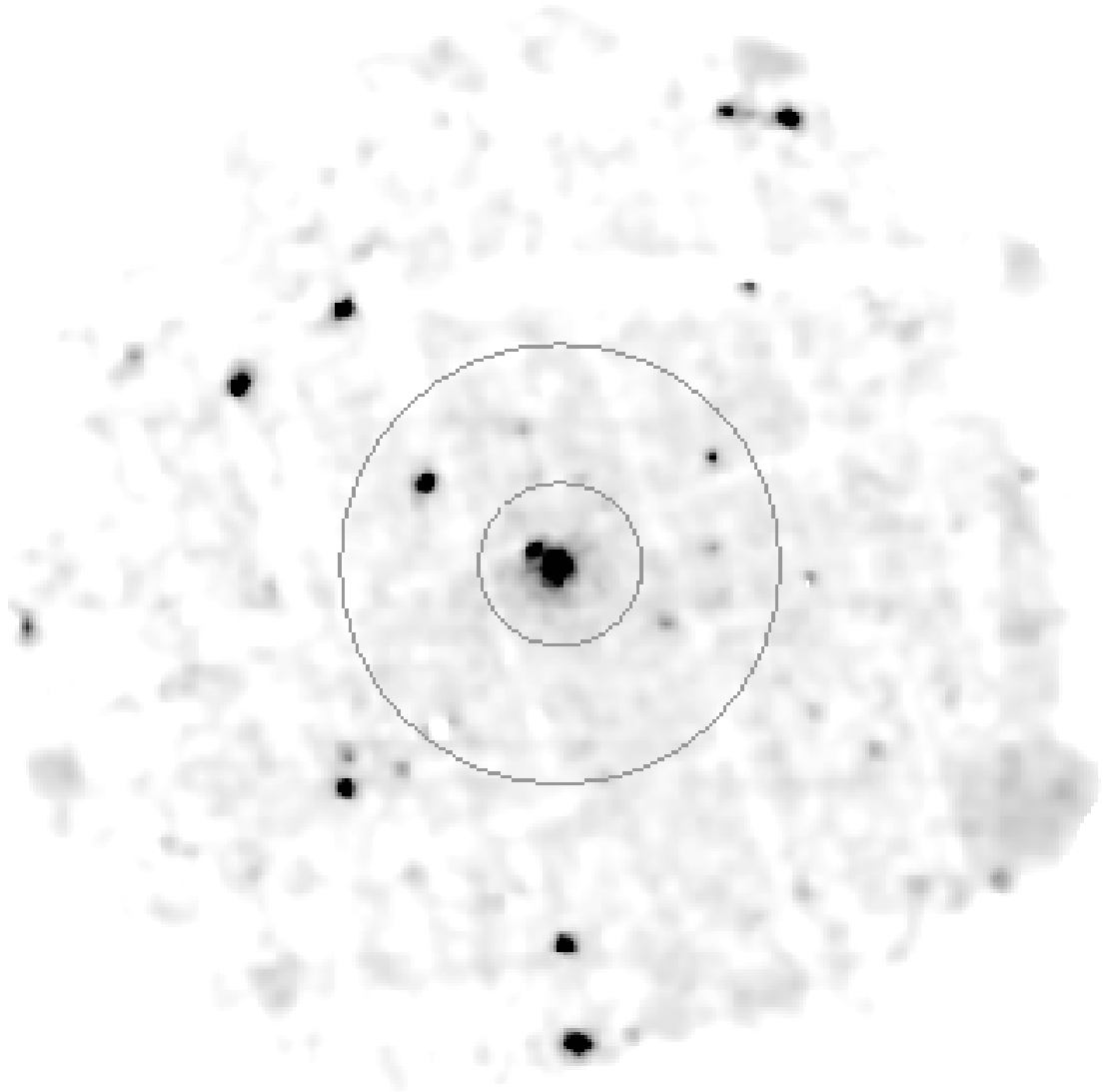}
 \caption{X-ray images for NGC 4382 (upper) and NGC 5363 (lower) in 
0.3-7 keV band. Data from all three EPIC cameras have been combined, particle 
background subtracted, exposure corrected and adaptively smoothed to form 
these images. The images have a field-of-view diameter of 30 arcmins. 
The pixel size is 4.4 arcseconds. North is up and east is left. 
The greyscale covers 5 counts/pixel above background and the location of 
the background annulus (11 arcminute diameter) is shown in each case. 
The inner circle is the D25 diameter.}
\end{figure}

\subsection{XMM-Newton observations}
NGC 4283 was observed on 2004 July 7 (ObsIDs 0201670101) with 
observation time of 36057 sec and NGC 5363 was observed on 2004 July 27 
(ObsID 0201670201) with observation time of 40278 sec. 
The observations were taken
in full frame mode with the medium optical blocking filter to reduce
unwanted background. Data for all three EPIC cameras was analysed
simultaneously for each galaxy. The spectral range covered by EPIC is 0.15
to 15 keV. See Jansen et al. (2001) for details of the XMM-Newton mission
and hardware.

The raw observation data files (ODFs) were retrieved from the XMM-Newton 
Science Archive. Calibration files were selected from the master 
calibration database, with the observation date of the target and an analysis 
date of 2005 April 20. The script {\it cifbuild} on the XMM-Newton website 
was used to select the appropriate calibration files for downloading.
XMM-Newton Science Analysis System (SAS) software was used to calibrate 
and analyse the event datasets and to create products including spectra, 
images and time series. SAS version 6.1 (release 
xmmsas\_20041122\_1834-6.1.0) was installed on a linux pc system, together 
with the FTOOLS and XSPEC version 12 software packages. Environment 
variables were set up to locate the calibration and ODF files, with a 
summary file in the working directory for each target. Event lists were 
generated using {\it emproc} and {\it epproc}.

A time series of events from each camera was generated using the graphical
interface {\it xmmselect} to the events selector routine {\it eveselect} in
the SAS software. The time series revealed that about half the observation
time was affected by X-ray background flares. These times were eliminated.
Further cleaning of the data was done using the three sigma clipping
routine of Ben Maughan (http://www.sr.bham.ac.uk/xmm2/scripts.html), on
data in the energy range 2 - 15 keV. This energy range was chosen because
the flaring was apparent at high and intermediate energies. X-ray events
were accepted for patterns 0-12 for the two EPIC MOS cameras and 0-4 for
the EPIC PN camera, and the events lists were further filtered using 
{\it FLAG=0}.

Inspection of the X-ray images (see Fig.~1) revealed that the optical 
isophotal diameter D25 (see Table~1) encompassed most of the X-ray 
counts from each galaxy. Therefore an annulus just outside this 
region was chosen to determine the background spectrum, extending out to 
a diameter of 11 arcminutes. Regions centred on a few bright point 
sources were removed from the background annulus in each case, before 
creating spectra. Analysis 
of out-of-field events revealed that the background was still 
dominated by particle events, therefore no vignetting correction was 
made for the background. The routine {\it emchain} was used to sort 
out-of-field events in the target data and also in long exposure, closed 
event files produced by Phillipe Marty, obtained from the web at 
http://www.sr.bham.ac.uk/xmm3. The script {\it CompareoutofFOV}, 
also available from the above Birmingham University web site, 
was used to calculate the scaling between the particle levels in the 
target data and closed data in the out-of-field region.

Source region and background region spectra were created from the 
cleaned photon events. 
A summary of the useful exposures and source counts obtained is given in 
Table~2. The percentages show that the source counts within D25 diameter
are about the same as the estimated background counts in that region, 
for each galaxy.

\bigskip
\begin{table*}
\centering
\begin{minipage}{140mm}
\caption{\bf XMM-Newton observations for two young, early-type galaxies. 
The percentage of counts in the specified region that are source counts 
is given in brackets.}
 
\begin{tabular}{lllll}
 & & & &  \\
{\bf Galaxy} &{\bf EPIC} &{\bf Useful} &{\bf Source counts} &{\bf Source counts} \\
       &{\bf camera} &{\bf exp.(s)} &{\bf in D25 diameter} &{\bf in 4r$_e$ diameter} \\
 & & & &  \\
NGC 4382  &MOS1  &17830 &2002$\pm74$ (50.7\%) &1382$\pm45$ (72.8\%) \\
NGC 4382  &MOS2  &17850 &2001$\pm73$ (51.1\%) &1302$\pm44$ (72.2\%) \\
NGC 4382  &PN    &12400 &4992$\pm123$ (53.2\%) &3447$\pm82$ (73.2\%) \\
 & & & &  \\
NGC 5363  &MOS1  &22960 &1360$\pm54$ (49.8\%) &789$\pm31$ (86.8\%) \\
NGC 5363  &MOS2  &22960 &1346$\pm54$ (50.2\%) &730$\pm32$ (86.5\%) \\
NGC 5363  &PN    &10620 &2145$\pm83$ (45.8\%) &1351$\pm59$ (84.7\%) \\
 & & & &  \\
\end{tabular}
\end{minipage}
\end{table*}
 
Response files were created for source regions and area scaling was
calculated using the SAS subroutines {\it arfgen}, {\it rmfgen} and 
{\it backscale}. The FTOOLS subroutine {\it grppha} was used to group the
source spectra into a minimum of 20 counts per bin. Data below 0.3 keV are
strongly affected by the soft photon background and by imperfections in the
calibration, and there are few source counts above about 7 keV. Also,
several background fluorescence lines occur above about 7 keV. Therefore
data within 0.3 to 7 keV was analysed for source spectral properties. One
or two weak background fluorescence emission lines remain around 1.5 keV,
but these did not significantly affect the spectral fits.

\begin{figure}
\includegraphics[angle=0,width=85mm,trim=0 0 0 0]{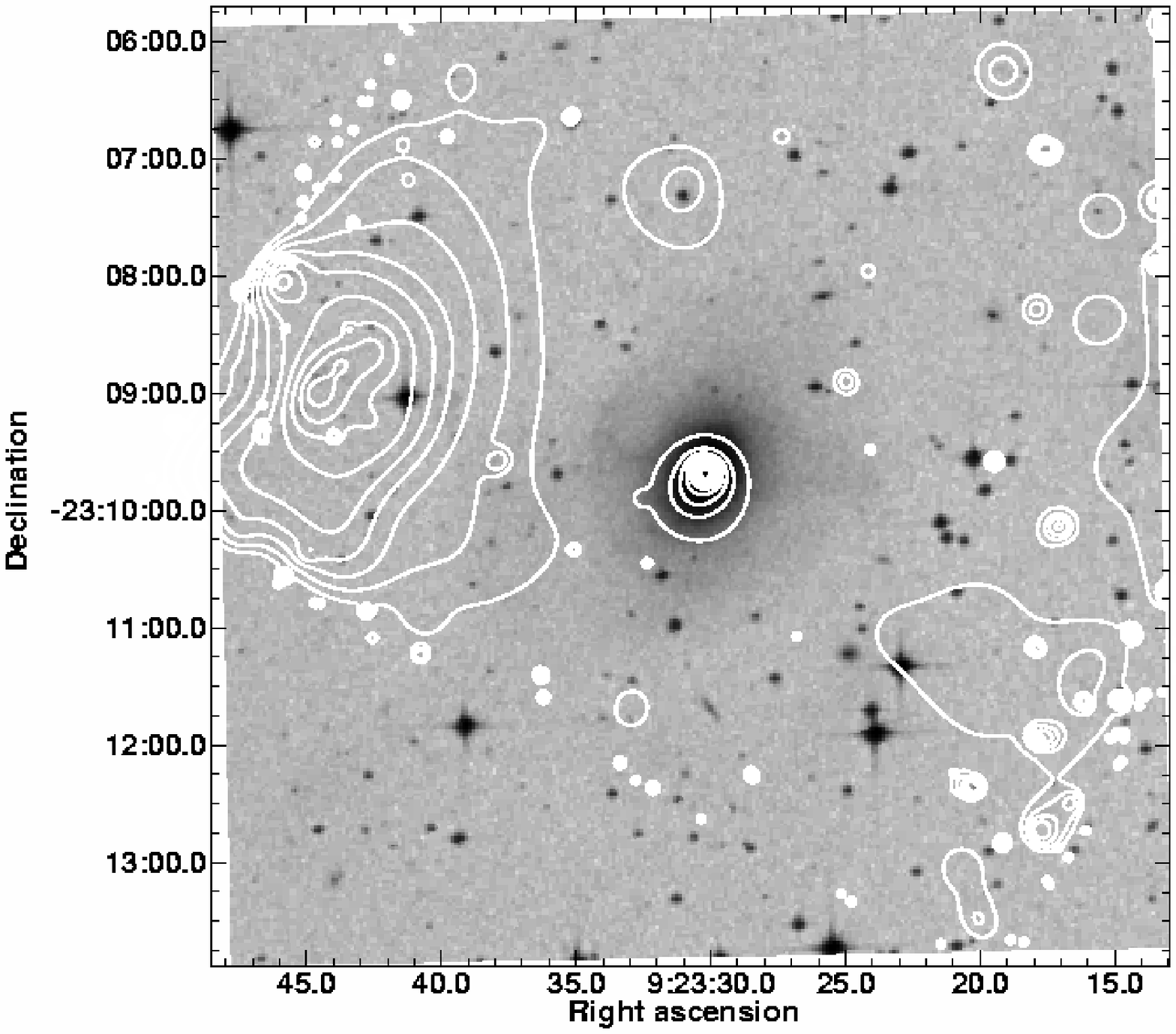}
\includegraphics[angle=0,width=85mm,trim=0 0 0 0]{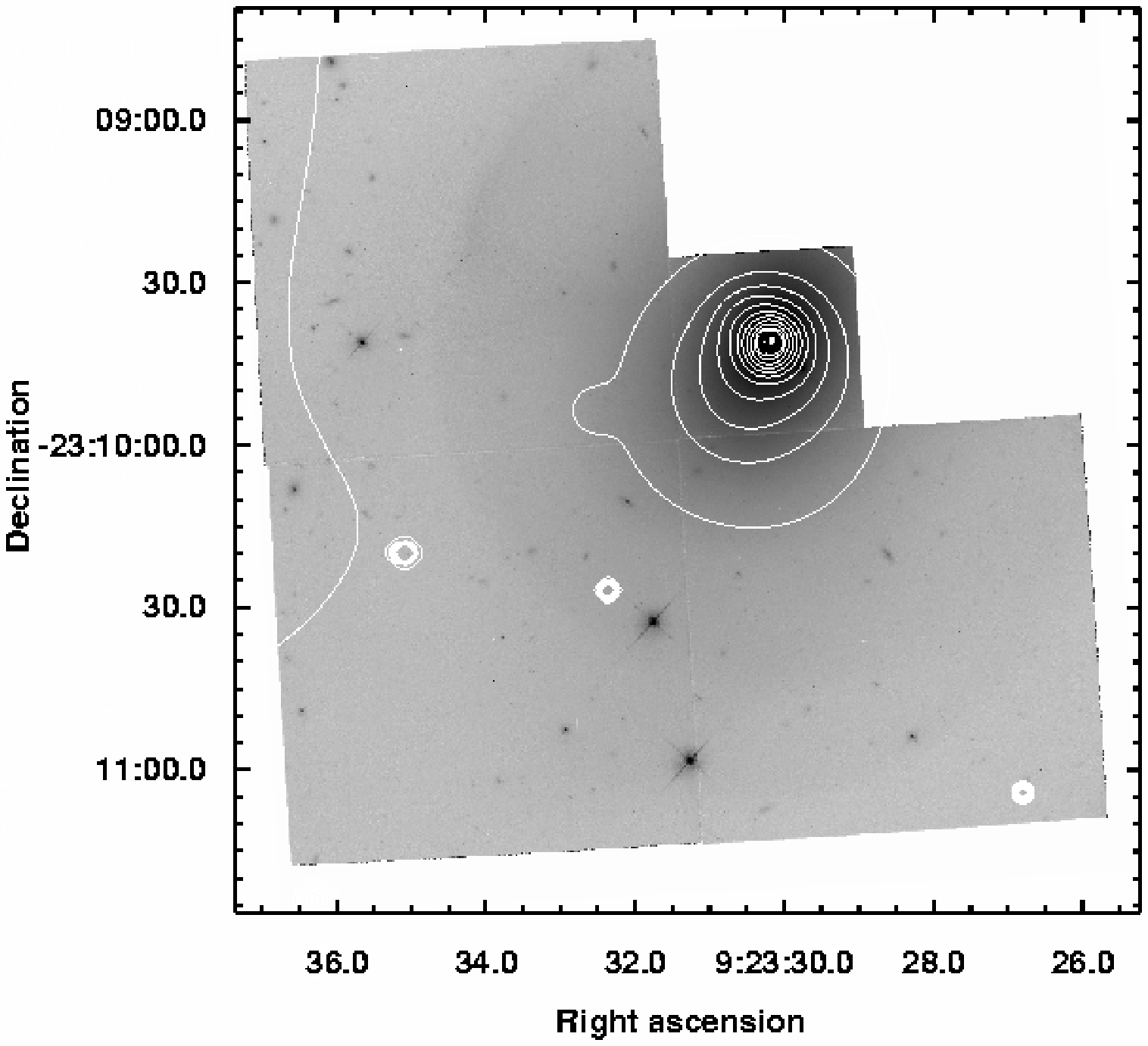}
 \caption{Upper plot: X-ray contours from Chandra observations of NGC 2865, 
overlaid on a DSS image. A close-up of the X-ray contours are shown over 
an HST image in the lower plot.}
\end{figure}

\subsection{Chandra Observations}

NGC~2865 was observed with the ACIS-S detector on the Chandra
observatory for a total of 29.9 ksec using the nominal aimpoint for 
the spectroscopy array. The observation date was 2001 May 17 (ObsID 2020).
A digitised sky survey (DSS) image is shown with X-ray contours
overlaid in Fig.~2. A close-up of the X-ray contours over an HST image is 
also shown in Fig.~2. NGC 2865 is clearly detected, with point source 
emission near the centre, surrounded by more diffuse emission associated with
NGC~2865.

NGC~4382 was observed for $\sim$40 ksec on 2001 Jan 29 (ObsID 2016) with
Chandra ASCI-S, in faint mode. The raw data was reprocessed with
\textsc{ciao} v3.1, and filtered to exclude bad pixels and events with ASCA
grades 1, 5 and 7. Filtering for periods of high background and to remove
point sources was carried out as described in O'Sullivan \& Ponman (2004a).
Our purpose in analysing the dataset was to produce a high resolution
surface brightness image, and a background image was extracted from the
blank-sky data described by
Markevitch\footnote{http://asc.harvard.edu/cal/}. The background image was
scaled to match the data by comparison of the counts in PHA channels
2500-3000.

\section[]{Derived X-ray properties}

\subsection{XMM-Newton spectra}
Spectral fitting was carried out using XSPEC, including thermal (MEKAL) 
and power-law components in the models. These were multiplied by
an absorption model (wabs). Absorption was assumed to be at least that 
due to our Galaxy. The Galactic HI absorption along the line-of-sight to 
each target galaxy was estimated using the script available at 
http://heasarc.gsfc.nasa.gov/cgi-bin/Tools/w3nh/w3nh.pl, based on data 
reviewed in Dicky \& Lockman (1990). Attempts to fit the absorption 
column produced uncertain results, consistent with zero absorption. 
Since we know that there is absorption through our Galaxy, we generally 
fix this in the absorption component. The results of one and two component 
fits are given in Tables~3 and 4 for NGC 4382 and NGC 5363 respectively.
Best fit 2-component spectral models are shown in Fig.~3.

\begin{figure}
\includegraphics[angle=-90,width=90mm,trim=0 20 0 0]{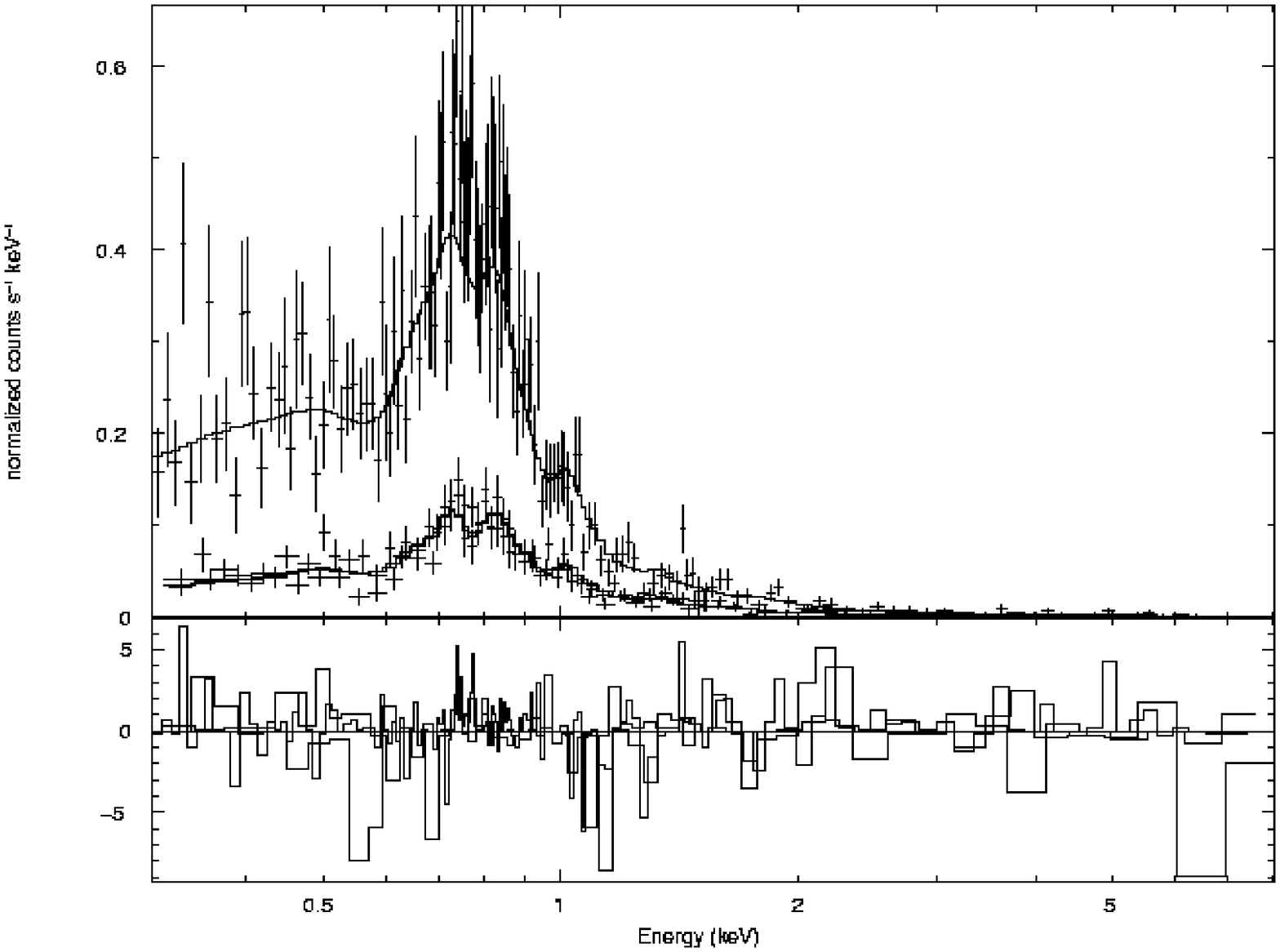}
\includegraphics[angle=-90,width=90mm,trim=0 20 0 0]{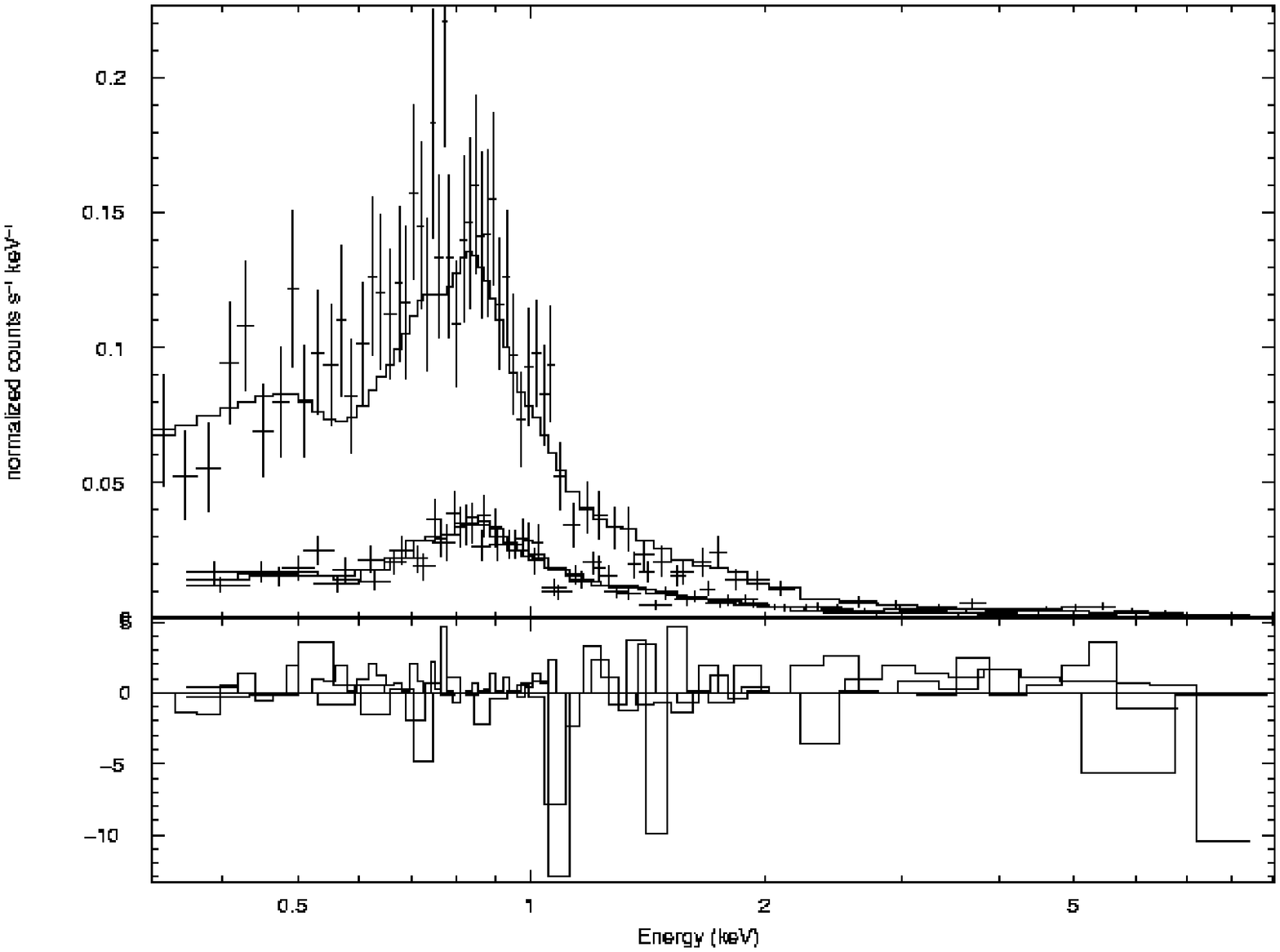}
 \caption{X-ray spectral fits to data within a diameter 4r$_e$,
for NGC 4382 (top) and NGC 5363 (lower). Normalised Counts s$^{-1}$ 
keV$^{-1}$ are shown against Energy in keV. Chi-squared plots are also shown 
in each case, highlighting any regions of disagreement between data and 
models.}
\end{figure}

\bigskip
\begin{table*}
\centering
\begin{minipage}{140mm}
\caption{\bf Spectral fits for NGC 4382, from XMM-Newton observations.
Columns are: hydrogen column (nH), temperature of the thermal component 
(kT$_1$), abundance of the thermal component relative to solar (Ab$_1$), 
flux in the thermal component (Flux$_1$), index of the power-law 
component (PL). The reduced chi-squared ($\chi^2_{\nu}$) and degrees of 
freedom (dof) are shown for each fit. Observed (unabsorbed) fluxes are 
in units of $10^{-13}$erg cm$^{-2}$s$^{-1}$ and errors are $\pm 1\sigma$ 
allowing for one interesting parameter, except where otherwise specified. 
Where ``Two interesting parameters:'' is specified this means that these 
two parameters were allowed to vary freely in the error determinations, 
together with the normalisations of the two components. Where no errors 
are given these parameters are fixed. ``Gal'' indicates the column density 
through our Galaxy along the line-of-sight to NGC 4382. The redshift is 
0.00243 from NED. See Section 3.1 for a discussion of these fits.}

\begin{tabular}{cccccccc}

 & & & & & & & \\
\multicolumn {8}{l}{\bf Within 4r$_e$ diameter} \\
{\bf nH} &{\bf kT$_1$} &{\bf Ab$_1$} &{\bf Flux$_1$} &{\bf PL} 
&{\bf $\chi_{\nu}^2$} &{\bf dof} &{\bf Total Flux} \\
{\bf $10^{22}$cm$^{-2}$} &{\bf keV} & {\bf Solar} &{\bf 0.3-7keV} 
&{\bf index} & & &{\bf 0.3-7keV} \\
\multicolumn {8}{l}{\it Single component model} \\
0.0251=Gal  & 0.379$\pm$0.007  & 1.0  & &  & 3.37   & 280   & \\
0.00$\pm$0.028 & 0.457$\pm$0.012  & 1.0  & &  & 3.49   & 279   & \\
0.0251=Gal & 0.519$\pm$0.012  & 0.082$\pm$0.006 & &  & 2.04   & 279   & \\
\multicolumn {8}{l}{\it Two component model} \\
0.0251=Gal  & 0.389$\pm$0.013  & 1.0  & 2.07 & 2.10$\pm$0.07  & 1.21  & 278  & 5.03 \\
0.0251=Gal & 0.409$\pm$0.015  & 0.151$\pm$0.029 & 3.05  & 1.47$\pm$0.15  & 1.15  & 277  & 5.29 \\
\multicolumn{2}{r}{\bf Two interesting parameters:} & 0.15 to $>$3 &  & 1.4 to 2.2 & \multicolumn{3}{l}{($\Delta\chi^2<2.30$, i.e. 1$\sigma$ range)} \\
\multicolumn {8}{l}{\it Three component model} \\
0.0251=Gal  & 0.406$\pm$0.016  & 1.0  &  & 1.74$\pm$0.10  & 1.16  & 276  &  \\
            & $<0.08$          & 1.0  &  &                &       &      &  \\
 & & & & & & & \\
\multicolumn {8}{l}{\bf Within D25 diameter} \\
{\bf nH} &{\bf kT$_1$} &{\bf Ab$_1$} &{\bf Flux$_1$} &{\bf PL} 
&{\bf $\chi_{\nu}^2$} &{\bf dof} &{\bf Total Flux} \\
{\bf $10^{22}$cm$^{-2}$} &{\bf keV} & {\bf Solar} &{\bf 0.3-7keV} 
&{\bf index} & & &{\bf 0.3-7keV} \\
\multicolumn {8}{l}{\it Single component model} \\
0.0251=Gal  & 0.368$\pm$0.007  & 1.0  &  &  & 2.26   & 488   & \\
0.00$\pm$0.04 & 0.382$\pm$0.014  & 1.0  &  &  & 2.23   & 487   & \\
0.0251=Gal & 0.489$\pm$0.012  & 0.075$\pm$0.005 &  &  & 1.53   & 487   & \\
\multicolumn {8}{l}{\it Two component model} \\
0.0251=Gal  & 0.389$\pm$0.014  & 1.0 & 3.03 & 2.22$\pm$0.07  & 1.19  & 486  & 7.77 \\
0.0251=Gal & 0.411$\pm$0.016  & 0.109$\pm$0.015 & 5.18 & 1.18$\pm$0.17  & 1.14  & 485  & 8.66 \\
\multicolumn {8}{l}{\it Three component model} \\
0.0251=Gal  & 0.398$\pm$0.015  & 1.0 &      & 1.68$\pm$0.12  & 1.16  & 484  &\\
            & $<0.08$          & 1.0 &      &                &       &      &\\
 & & & & & & & \\
\end{tabular}
\end{minipage}
\end{table*}

\bigskip
\begin{table*}
\centering
\begin{minipage}{140mm}
\caption{\bf Spectral fits for NGC 5363, from XMM-Newton observations.
Columns are as in Table.~3. Observed (unabsorbed) fluxes are in 
units of $10^{-13}$erg cm$^{-2}$s$^{-1}$ and errors are 
$\pm 1\sigma$. ``Gal'' indicates the column density through our Galaxy
along the line-of-sight to NGC 5363. The redshift is 0.00380 from NED.
See Section 3.1 for a discussion of these fits.}

\begin{tabular}{cccccccc}

 & & & & & & & \\
\multicolumn {8}{l}{\bf Within 4r$_e$ diameter} \\
{\bf nH} &{\bf kT$_1$} &{\bf Ab$_1$} &{\bf Flux$_1$} &{\bf PL} 
&{\bf $\chi_{\nu}^2$} &{\bf dof} &{\bf Total Flux} \\
{\bf $10^{22}$cm$^{-2}$} &{\bf keV} & {\bf Solar} &{\bf 0.3-7keV} 
&{\bf index} & & &{\bf 0.3-7keV} \\
\multicolumn {8}{l}{\it Single component model} \\
0.0208=Gal  & 0.969$\pm$0.021  & 1.0  & &  & 7.45   & 137   & \\
0.000$\pm$0.037 & 0.982$\pm$0.021  & 1.0  & &  & 7.35   & 136   & \\
0.0208=Gal & 1.053$\pm$0.041  & 0.044$\pm$0.009 & &  & 3.54   & 136   & \\
\multicolumn {8}{l}{\it Two component model} \\
0.0208=Gal  & 0.603$\pm$0.024  & 1.0  & 0.51 & 1.69$\pm$0.05  & 1.30  & 135  & 2.55 \\
0.0208=Gal & 0.609$\pm$0.025  & 0.171$\pm$0.059 & 0.84 & 1.43$\pm$0.13  & 1.28  & 134  & 2.57 \\
\multicolumn {8}{l}{\it Three component model} \\
0.0208=Gal & 0.660$\pm$0.043  & 1.0  &    & 1.52$\pm$0.07  & 1.23  & 133  &\\
           & 0.244$\pm$0.081  & 1.0  &    &                &       &      &\\
 & & & & & & & \\
\multicolumn {8}{l}{\bf Within D25 diameter} \\
{\bf nH} &{\bf kT$_1$} &{\bf Ab$_1$} &{\bf Flux$_1$} &{\bf PL} 
&{\bf $\chi_{\nu}^2$} &{\bf dof} &{\bf Total Flux} \\
{\bf $10^{22}$cm$^{-2}$} &{\bf keV} & {\bf Solar} &{\bf 0.3-7keV} 
&{\bf index} & & &{\bf 0.3-7keV} \\
\multicolumn {8}{l}{\it Single component model} \\
0.0208=Gal  & 0.995$\pm$0.019  & 1.0  &  &  & 3.88   & 330   & \\
0.000$\pm$0.016 & 5.76$\pm$0.434  & 1.0  &  &  & 2.05   & 329   & \\
0.0208=Gal & 1.75$\pm$0.011  & 0.000$\pm$0.039 &  &  & 1.70   & 329   & \\
\multicolumn {8}{l}{\it Two component model} \\
0.0208=Gal  & 0.610$\pm$0.031  & 1.0 & 0.69 & 1.79$\pm$0.05  & 1.18  & 328  & 4.37 \\
0.0208=Gal & 0.629$\pm$0.028  & 0.065$\pm$0.015 & 1.85 & 1.06$\pm$0.15  & 1.12  & 327  & 4.78 \\
\multicolumn {8}{l}{\it Three component model} \\
0.0208=Gal  & 0.670$\pm$0.034  & 1.0 &    & 1.42$\pm$0.08  & 1.08  & 326  & \\
            & 0.198$\pm$0.025  & 1.0 &    &                &       &      & \\
 & & & & & & & \\
\end{tabular}
\end{minipage}
\end{table*}

To check our simple background subtraction technique we also tried the
double background subtraction method (described by Arnaud et al. 2002).
The background spectrum thus generated looked very similar to our simple
background spectrum, but with more noise at high energies.  The $\sim$ 17\%
vignetting of photon events expected in the background annulus is a much
smaller fraction of the total background, since the background was still
dominated by particles even after cleaning and clipping.  Trials with this
double subtraction led to very similar thermal component parameters, but
noisy power-law parameters.  Therefore, given the uncertainties in the
background we retain our analysis with the simpler background estimate.
This samples the X-ray background from the same time as the source was
observed, minimising temporal changes.

Tables~3 and 4 show the following results for the spectral fits:
\begin{enumerate}
\item Single component models (wabs$\times$mekal or wabs$\times$powerlaw) 
  are ruled out for 
  both NGC 4382 and NGC 5363, from the high reduced chi-squareds and 
  from visual inspection of the systematic residuals.
\item Two component models (wabs(mekal+powerlaw)) can almost fit the data
  (reduced chi-squareds ~1.1 to 1.3). Adding another mekal model does
  not lead to significantly better fits. 
\item Residuals in the two component fits appear 
  at specific energies (between $\sim$1 and 2 keV, and at $>$5 keV - 
  see Fig.~3). These are probably due to small residuals from fluorescence 
  lines generated in the EPIC cameras, described in the XMM-Newton Users 
  Handbook.
\item The column density (nH) is fixed at the Galactic values, since attempts
  to fit this resulted in zero column density, which is physically
  unrealistic.
\item The temperature of the mekal component is well constrained, especially
  in NGC 4382, even when assumptions are changed about the metallicity of 
  the hot gas component.
\item The index of the power-law and the abundance in the mekal model can
  mimic similar fits to the low energy data. Therefore these two parameters
  are not individually well constrained. This is illustrated by the results
  of stepping through these two parameters in the case of NGC 4382, for
  data within a diameter of 4r$_e$: adequate fits include abundance 
  Ab$_1$=0.15 (relative to solar) and power-law index PL=1.4, 
  through to Ab$_1$=3 (or greater) and PL=2.2. These results are shown 
  for NGC 4382 under the heading of ``Two interesting parameters:'' in 
  Table.~3. The range of acceptable fits is given, allowing for two 
  interesting parameters and thus $\Delta \chi^2$ of $<2.3$ above the minimum. 
  The power-law index is
  separately constrained to be PL$\sim$1.7 from visual inspection of the 
  high energy data in the spectrum, which shows systematic deviations 
  from the model for poor fits to the power-law component.
\item The overall proportions of flux in the hot gas component, after 
  correcting for Galactic absorption, are: 0.60 for NGC 4382 and 0.39 for 
  NGC 5363. Therefore NGC 4382 contains proportionally more gas than NGC 5363, 
  which is dominated by the power-law component describing stellar 
  contributions (e.g. Matsushita et al. 1994). Therefore there is little 
  evidence of large quantities of hot gas in NGC 5363. 
\end{enumerate}

So, in summary these data need at least two components and the MEKAL 
temperature is quite well constrained. To estimate the hot gas mass associated
with the MEKAL component we need to measure radial temperature changes and
deproject the observed X-ray brightness profile. Of the three galaxies 
analysed here this is only possible for NGC 4382, since that observation 
has enough counts to do so and it has a large proportion of its X-ray flux 
in the MEKAL component. Thus in Section 4 we aim to use the X-ray 
properties to estimate the gas and overall mass distribution 
in NGC 4382. Table~5 summarises the best 2-component fits for our target 
galaxies.

\subsection{Chandra spectrum}
Examining the lightcurve from chip 7 (the back illuminated CCD) we find 
only a small enhancement of the count rate so we analyse the entire 
observation here.  The background was obtained from a circle just outside 
the D25 radius near the centre of the chip. We also used the blank 
sky backgrounds to fit the spectra and found no significant difference in
the fits.  All detectable point sources were excluded from both source
and background spectra, using a region size appropriate for the
off-axis angle of the source.  Finally, the extracted spectra were
rebinned so that each channel has a minimum of 25 counts. We also
investigated the effects of using background spectra extracted from
regions further from the galaxy centre, which reduces any possible
contamination of the background from the galaxy, but increases the
vignetting correction; no significant difference was found in the
results.

Data was extracted from a diameter of 8r$_e$, excluding the central 2 arcsecs 
radius (95\% of the encircled point source energy) so that the diffuse 
gas is not
contaminated by the central point source. A pure MEKAL model with Galactic
absorption can be rejected with a reduced chi-squared of 1.88 for 23 dof.
The pure power-law model can be rejected with a reduced chi-squared of 1.45
for 24 dof. A model with wabs*(mekal + brems) yields a best fit temperature
for the MEKAL component of 0.32$^{+0.10}_{-0.04}$ keV with 90\% confidence, 
with the abundance only constrained to be above 0.14 solar. The reduced 
chi-squared is 0.86 for 22 dof. The higher energy component was fixed at
7.3 keV bremsstrulung emission, representing LMXBs (Irwin, Athey \& Bregman 
2003). The total absorbed flux is $6.44\times 10^{-14}$ ergs cm$^{-2}$ 
s$^{-1}$ in the 0.3 to 7 keV range (unabsorbed flux is $7.47\times 10^{-14}$ 
ergs cm$^{-2}$ s$^{-1}$). The absorption was fixed at the Galactic value 
(nH=6.5$\times 10^{20}$) and there are 687 source counts.

The spectrum for the central source in NGC~2865 was extracted using an
aperture that encloses 95\% of the encircled energy (2 arcsec radius) 
and the spectrum
was binned so that each channel has a minimum of 20 counts. The
background region is far enough from the galaxy so that no counts from
the diffuse emission are included.  Since the extraction region for the
central source contains diffuse emission from the galaxy we fit the
point source with a power-law plus a thermal plasma model. Since we
only have 127 counts in this spectrum we fix the parameters of thermal
plasma model to the value found above and only allow its normalisation
to vary. The power-law components are then fitted resulting in a
power-law index of 1.67$^{+1.94}_{-1.33}$ with $\chi^2_\nu$ = 0.21 for
3 dof. The absorbed flux from the power-law component is $1.69\times 10^{-14}$ 
ergs cm$^{-2}$ s$^{-1}$ (unabsorbed flux is $2.07\times$10$^{-14}$ 
ergs s$^{-1}$ cm$^{-2}$).

There is also a region of more extended diffuse X-ray emission to the left of,
and separate from NGC 2865, which has no clear optical counterpart in the DSS 
image. It was first found as an unidentified X-ray source in the ROSAT All Sky 
Survey, Bright Source Catalogue (Voges et al. 1999) and is called
1RXS J092344.1-230858. The source is clearly extended and appears to have 
a double peaked core. The spectrum was extracted using an elliptical region 
centred at 09:23:43 -23:08:51.64 (J2000) with a major axis of 
115 arcsec and 75 arcsec minor axis with PA. 145 deg. This is slightly 
offset from the peak of the extended diffuse emission 
(9:23:44.5, -23:08:59.5) to 
avoid the edge of the chip.  The spectrum was fit with an absorbed thermal 
spectrum (wabs*apec) with wabs column fixed at nH=$6.5\times 10^{20}$ 
cm$^{-2}$ and the abundance fixed to 0.3 solar. The temperature and the 
redshift were then fitted yielding kT= 6.4$^{+1.6}_{-1.2}$ with a redshift of 
0.2335$^{+0.0708}_{-0.023}$. The goodness of fit was 0.65 for 165 dof. 
The luminosity is $1.9\times 10^{44}$ ergs s$^{-1}$ for the redshift given 
above. Given the luminosity and extended nature of the source this is most 
likely a galaxy cluster. This offset diffuse emission is not included in our 
assessment of the X-ray emission from NGC 2865.

Summing the central and diffuse emission from NGC 2865 gives an overall
absorbed flux of $8.47\times 10^{-14}$ erg s$^{-1}$ cm$^{-2}$ (unabsorbed 
flux of $1.00\times 10^{-13}$ erg s$^{-1}$), including 
the diffuse flux in the central region.
NGC 2865 is quite a weak X-ray source in comparison to some other 
ellipticals, as will be shown in Section 5.

\bigskip
\begin{table*}
\centering
\begin{minipage}{140mm}
\caption{\bf Overall best fitting 2-component model parameters for the three 
galaxies analysed in this paper. 90\% errors are given for 1 interesting 
parameter. Fixed parameters are indicated without errors.}

\begin{tabular}{llllll}
 & & & & & \\
{\bf Galaxy} &{\bf nH} &{\bf kT$_1$} &{\bf Ab$_1$} &{\bf PL index} & $\chi_{\nu}^2$ \\
       &{\bf (10$^{22}$ cm$^{-2}$} &{\bf (keV)} &{\bf (Solar)} & & \\
 & & & & & \\
NGC 4382  &0.0251  &0.411$^{+0.026}_{-0.023}$ &0.109$^{+0.040}_{-0.009}$ 
&1.18$^{+0.24}_{-0.16}$ & 1.14 \\
NGC 5363  &0.0208  &0.629$^{0.046}_{-0.046}$ &0.065$^{+0.023}_{-0.016}$ 
&1.06$^{+0.13}_{-0.12}$ & 1.12 \\
NGC 2865  &0.0650  &0.320$^{+0.10}_{-0.04}$ & $>0.14$ &7.3 keV brems & 0.86 \\
 & & & & & \\
\end{tabular}
\end{minipage}
\end{table*}

\section[]{Mass distribution in NGC 4382}

Under the assumption of hydrostatic equilibrium, it is possible to estimate
the distribution of mass and other properties (entropy, cooling time, etc.)
based on the temperature and density of the gaseous halo. We therefore
extracted spectra from four radial bins and fitted them to determine the
temperature distribution, and fitted the radial surface brightness
profile, from which gas density can be calculated.  It is difficult to
be certain whether the gas is in fact in hydrostatic equilibrium, but the
X-ray image of NGC~4382 appears to be indicate a relatively smooth
distribution, suggesting that its halo is undisturbed.

For the radial temperature profile, bins were chosen to have similar
numbers of source counts, slightly increasing in larger radial bins to
allow for the increased fraction of flux in the background. A bright point
source in the outermost radial range was removed for these thermal flux and
temperature determinations (see Fig.~1 top right in inner circle).  The
results of spectral fitting in 4 radial bins are shown in Table~6. These
are for wabs(mekal+powerlaw) 2-component fits, with nH=$2.51\times10^{20}$
cm$^{-2}$ (Galactic column) fixed and power-law index PL=1.7 fixed (from
inspection of the spectrum above 3 keV and previous fits). This value of
the power-law index describing the stellar contributions to the X-rays is
similar to that found from fits to ASCA data for elliptical galaxies
(PL=1.82$\pm$ 0.1, White et al. 2002) and from Chandra data for NGC 4382
specifically (1.52$\pm 0.11$, Sivakoff et al. 2003). Temperatures and
abundances are fitted and 1-sigma errors are given.\\

\bigskip
\begin{table*}
\centering
\begin{minipage}{140mm}
\caption{\bf Spatially resolved spectral fits to NGC 4382 XMM-Newton data.
Fluxes are $\times 10^{-13}$ erg s$^{-1}$ and are uncorrected for absorption, 
unless otherwise stated, in 0.3-7keV band.}
 
\begin{tabular}{lllllllll}
 & & & & & & & & \\
{\bf Radial} &{\bf Source} &{\bf \% of} &{\bf kT} &{\bf Ab} & {\bf $\chi^2_{\nu}$ (dof)}& {\bf Total} & {\bf MEKAL} & {\bf MEKAL flux} \\
{\bf range} &{\bf counts} &{\bf total} &{\bf (keV)} &{\bf (Solar units)} & & {\bf flux} & {\bf flux} & {\bf (unabsorbed)} \\
 & & & & & & & & \\
0 - 0.5r$_e$     &1539.9  &93 &0.495$\pm0.031$ &0.318$\pm0.124$ & 1.13 (73) & 1.24 & 0.464 & 0.53 \\
0.5r$_e$ - r$_e$    &1858.7  &85 &0.435$\pm0.028$ &0.134$\pm0.030$ & 1.07 (94) & 1.62 & 0.764 & 0.90 \\
r$_e$ - 2r$_e$      &2365.0  &65 &0.366$\pm0.018$ &0.173$\pm0.041$ & 1.26 (149) & 1.79 & 1.13 & 1.33 \\
2r$_e$ - 0.5D25  &2057.3  &34 &0.379$\pm0.032$ &0.111$\pm0.035$ & 1.09 (258) & 1.61 & 1.17 & 1.39 \\
 & & & & & & & & \\
\end{tabular}
\end{minipage}
\end{table*}


A radial temperature gradient was estimated from these results by fitting a
straight line through the temperature points. This gave:
$kT=-0.00948r+0.473$ keV (for radius $r$ in kpc). Fig.~4 shows a plot of
the temperature profile and linear fit.  A 2nd order polynomial fits
better, but is probably unrealistic at large radii, since the slight
increase in temperature in the outer bin (which is not statistically
significant) leads to an upturn in the polynomial model, which in turn
leads to an unphysical flattening of the mass profile.

\begin{figure}
\includegraphics[angle=0,width=90mm,trim=80 55 0 0]{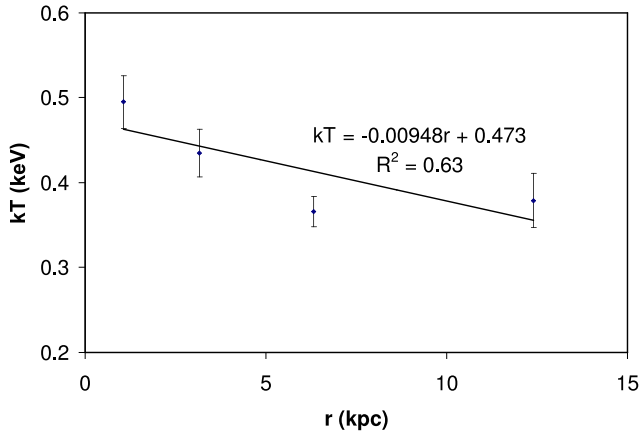}
 \caption{Measured temperatures and linear fit to the temperature profile
for NGC 4382 from XMM-Newton observation.}
\end{figure}

The unabsorbed, thermal component flux of NGC 4382 within D25 diameter is
estimated to be $4.15\times 10^{-13}$ erg cm$^{-2}$ s$^{-1}$ (assuming
PL=1.7).  For a distance of 15.9 Mpc the total unabsorbed luminosity of the
thermal component only is then $1.26\times 10^{40}$ erg s$^{-1}$. The
normalisation for the MEKAL component is: $7.41\times 10^{-4}\pm 1.56\times
10^{-4}$, with units as defined in the on-line XSPEC manual.

The X-ray surface brightness profile was estimated by simultaneous fitting
of our XMM-Newton data and the archival Chandra data for NGC 4382.  This
provides the useful combination of high spatial resolution in the core from
the Chandra data and high sensitivity from the XMM-Newton data, which
allows the outer, low surface brightness regions to be defined more
precisely.  The Chandra profile was based on an exposure corrected 0.3-2
keV ACIS-S3 image with point sources and background subtracted. For
XMM-Newton a 0.3-3 keV exposure corrected image was used, with point
sources subtracted but not background. This background level was instead
modelled out during fitting.  Circular annuli were used in both cases.  The
maximum diameter is D25 and minimum diameter is 2 arcsec based on Chandra
spatial resolution. PSF convolution was included in the XMM-Newton fit.
Appropriate PSF images for each camera were extracted from the calibration
database, the images were summed, and a radial profile taken. The fitting
was carried out in \textsc{ciao sherpa} using this profile as the PSF
model. PSF convolution was not included in the Chandra fit, as the Chandra
on-axis PSF is very narrow. Prior experience with similar datasets has
shown that PSF convolution has no significant effect on the fit. 

Initial fitting showed that a single beta model provided a poor
approximation to the data (reduced $\chi^2$ 3.27 for 85 dof, see
Fig.~5). Adding a second beta model produces an improved fit (see Fig.~5),
with reduced $\chi^2$ of 1.65 (82 dof). Addition of a central point
source is not favoured by the fit, demonstrating that the central component
is extended. The remaining residuals appear to be largely noise related,
rather than indicating the need for an additional component, so we have not
attempted fits with more complex models.  The best fit parameters for the
core radii ($r_c$) and $\beta$ values, with 1$\sigma$
errors, were: \\

\noindent $r_{c1}$ = 28.41\arcs$^{+4.75}_{-3.35}$\\
$\beta_1$ = 0.445$^{+0.010}_{-0.009}$ \\
$r_{c2}$ = 1.99\arcs$^{+1.15}_{-0.92}$ \\
$\beta_2$ = 0.602$^{+0.254}_{-0.143}$ \\

The outer component has a slightly smaller core and flatter $\beta$ than
the fit found by Sivakoff et al. (2003) for the Chandra data alone. Here the
fit is better constrained, mainly because of the XMM-Newton data at large
radii.  The addition of a second component also helps define the core of
the more extended component more accurately.  The fitted surface brightness
profile is shown in Fig.~5, incorporating two beta models and a constant
background level. 

\begin{figure}
\includegraphics[angle=0,width=90mm,trim=0 195 0 40]{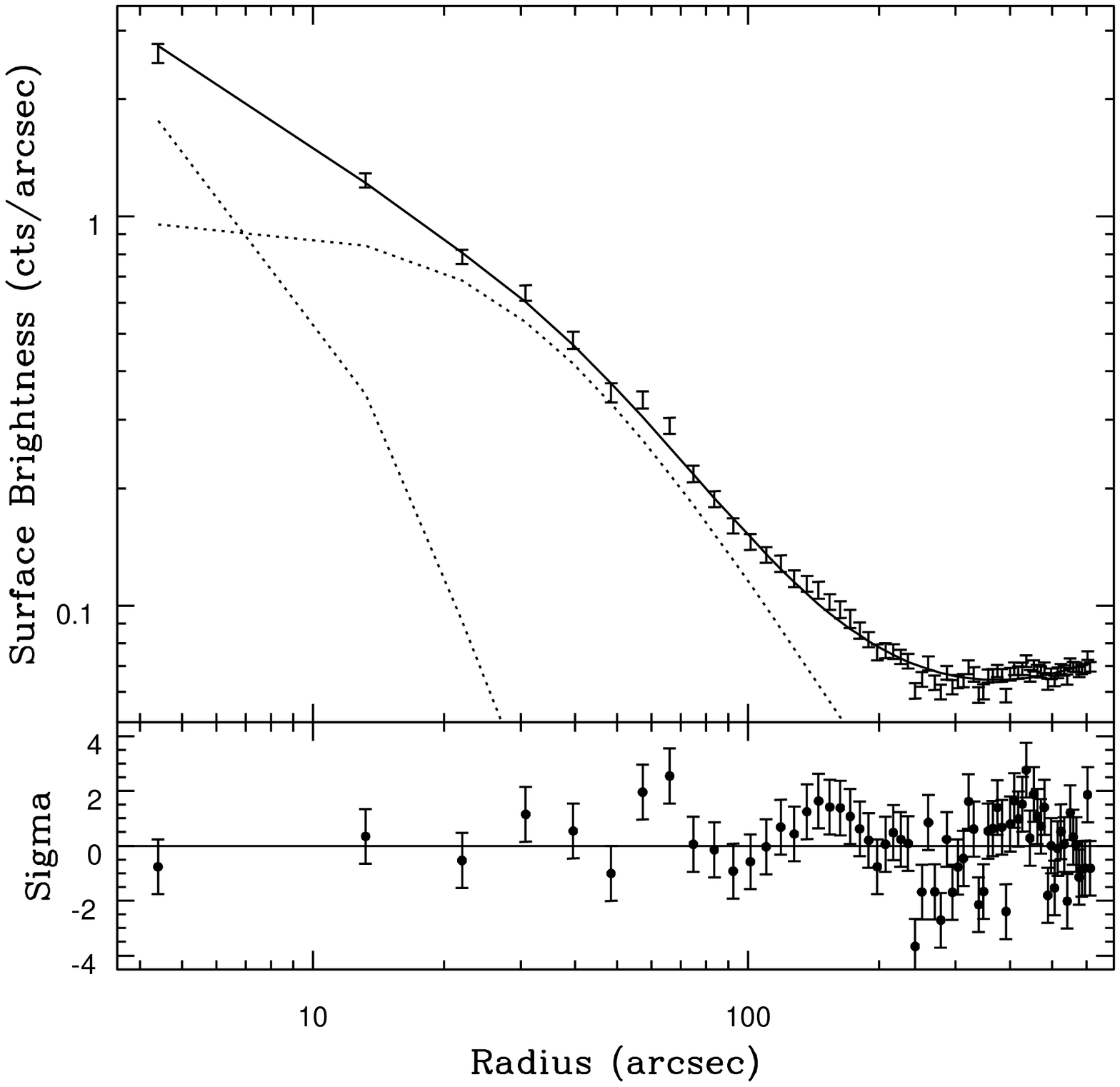}
\includegraphics[angle=0,width=90mm,trim=0 200 0 40]{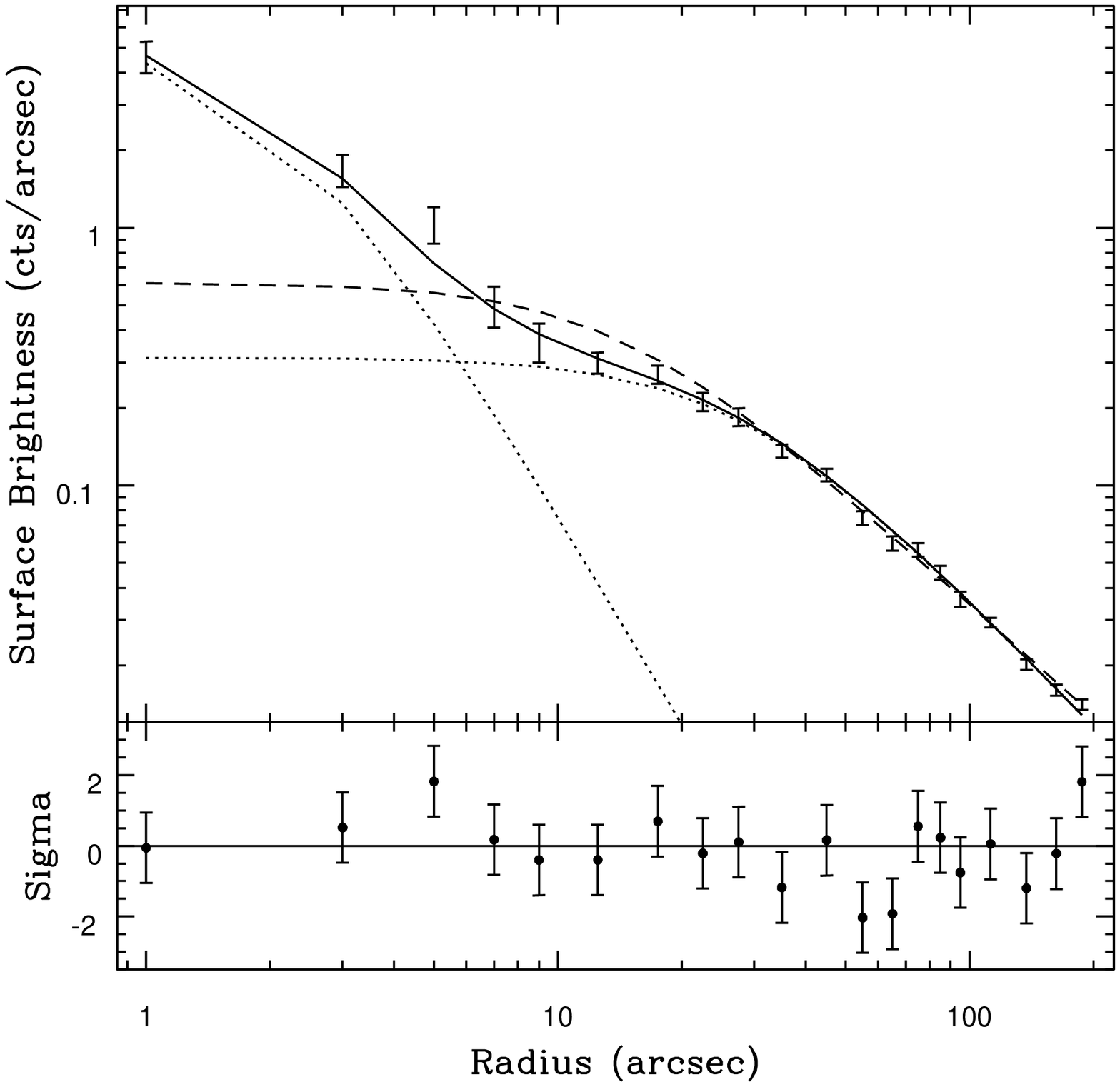}
 \caption{XMM-Newton 0.3-3.0 keV (\textit{Upper}) and Chandra 0.3-2.0 keV
   (\textit{Lower}) X-ray surface brightness profiles of NGC 4382, fitted
   by two beta models plus, for the XMM data only, a constant background.
   The best fitting model is marked as a solid line, the two beta models as
   dotted lines. Residuals from the fit are shown in terms of the
   significance of the deviation.  See Section 4 for the fitted
   parameters. The dashed line in the Chandra plot shows the best fit
   attained using a single beta model.}
\end{figure}

Our technique for calculating the mass profile based on these results is
described in O'Sullivan \& Ponman (2004b) and O'Sullivan et al. (2005).
Profiles of total mass, gas mass, gas entropy and cooling time are all
estimated from the temperature and density profiles.  The error on each
parameter is estimated through a monte-carlo process in which the measured
errors on temperature and surface brightness profiles, and other parameters
such as total luminosity, are used to vary the input parameters. The
mass-to-light ratio (M/L) is calculated by assuming the optical surface
brightness distribution is circular, follows a de Vaucouleurs profile, and
is normalised to match the B-band luminosity.

\begin{figure}
\includegraphics[angle=0,width=92mm,trim=10 60 10 50]{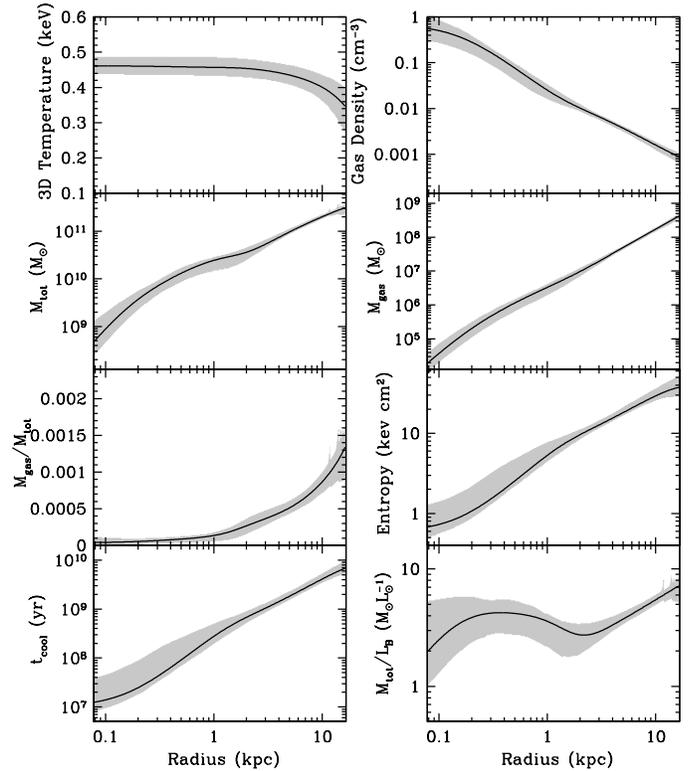}
 \caption{Reconstructed temperature, density, total and gas mass, gas
   fraction, entropy, cooling time and M/L ratio distributions for NGC
   4382, using the linear fit to the temperature profile from XMM-Newton
   data. The optical half-light radius is at $r_e$=4.2 kpc. This 
   reconstruction is discussed in Section 4.}
\end{figure}

Fig.~6 shows the reconstructed properties as a function of radius assuming
a linear temperature profile. We can compare NGC~4382 with NGC~4555, an
elliptical with an extensive hot gas halo, which is very isolated and
therefore likely to be relaxed and undisturbed.
(O'Sullivan \& Ponman 2004b). The gaseous halo of NGC~4382 is both cooler
and less dense than that of NGC~4555, and we find the gas mass within 10
kpc to be a factor of $\sim$4 lower than that of NGC~4555
($\sim$1.8$\times$10$^{8}$ M$_\odot$ compared to $\sim$7.7$\times$10$^{8}$
M$_\odot$). The total mass of NGC~4382 within the same radius
(2$\times10^{11}$ M$_\odot$) is also smaller by a similar factor, leading
to almost identical values of gas fraction, entropy and cooling time in the 
two systems, at a given radius. However, the mass-to-light ratio of NGC~4382 
is rather lower than that of NGC~4555, particularly within 10 kpc (5.4
M$_\odot$/L$_\odot$ compared to 14.7).  One might expect M/L$\sim$ 3 to 5
from stars alone (e.g. Sparke \& Gallagher 2000), and the M/L ratio in the
inner part of the galaxy is comparable to this, indicating that stars
dominate the mass within a $\sim 10$ kpc radius.  However, from Fig.~6
there appears to be additional gravitating mass further out from the centre
of NGC 4382 since the M/L ratio increases constantly beyond $\sim 5$ kpc
radius, reaching a maximum values of M/L$\sim$7.5 at 4r$_e$. We therefore 
expect that were we to be able to extend the mass
profile to larger radii, we would find a total NGC~4382 M/L ratio more
typical of elliptical galaxies, though perhaps lower than that of NGC~4555.

Recently, orbital modelling of planetary nebulae in some moderate
luminosity, early-type galaxies (NGC~821, NGC~3379, NGC~4494, Romanowsky et
al. 2003) has suggested that they have very low M/L ratios out to
$\sim$5Re. This may indicate that these ellipticals have either little dark
matter, or that their dark matter halos take a radically different form to
that predicted by the standard $\Lambda$CDM structure formation models.
However, the mass estimated for these galaxies is dependent on the choice
of orbital models, and these have been challenged (Dekel et al. 2005).
Simulations of merging spirals suggest that the orbits in the outer part of
the resulting post-merger elliptical are much more radial than was
previously expected. If such orbits are assumed when estimating the mass of
the Romanowsky ellipticals, a larger M/L ratio is found, consistent with a
normal dark matter halo. Modelling of the X-ray halo provides an
alternative method of measuring mass, and therefore in principle could be
used to resolve this issue. However, the three Romanowsky ellipticals are
all X-ray faint. In the case of NGC~821, its X-ray luminosity is so low
that no useful limit can be placed on its gas content (Fabbiano et al.
2004). Low gas content could be related to a number of factors; lack of a
dark matter halo could make it difficult for a galaxy to retain a gaseous
halo, but ram-pressure or tidal stripping, AGN activity or starburst driven
winds could also remove much of the gas. Wind activity following a
merger-induced starburst is the likely cause of the lack of hot gas in
post-merger galaxies. NGC~4382 has a comparably low X-ray luminosity, but
its optical luminosity (M$_B$=-21.01 mag) is nearly twice as bright as the
M$_B\sim -20.4$ mag galaxies in Romanowsky et al.  NGC 4382 also has a 
younger luminosity weighted age. The three intermediate luminosity 
ellipticals studied by Romanowsky et al. were not very young (with 
ages of 7.2, 9.3 and 6.7 Gyr for NGC 821, 3379 and 4494 respectively) 
from TF02. From these results it does not seem that optical spectroscopic 
youth is related to a lack of massive, dark-matter halo.
However, more systems and deeper observations are needed to check this, and
a number of projects are underway to investigate this issue.

\section{X-ray emission versus age}

In this section the data for the three galaxies is set in context with
published data, for galaxies with estimated ages. We discuss the benefits
and drawbacks of different age estimators. The motivation for looking at
age dependencies is to see how the gas evolves in early-type galaxies.
Ellipticals may be produced from mergers of spiral galaxies
(e.g. Bournaud, Jog \& Combes, 2005). The
observational evidence to support this idea is summarised by Schweizer
(1998).  Tidal features occur around elliptical galaxies that show other
evidence of youth in their optical spectra, colours and disturbed
morphology of their outer isophotes.  Using ROSAT data Fabbiano \&
Schweizer (1995) found that two such dynamically young ellipticals were X-ray
faint compared with other ellipticals. Later Mackie \& Fabbiano (1997, 32
galaxies) and Sansom, Hibbard \& Schweizer (2000, 38 galaxies) showed that
young early-type galaxies are X-ray faint generally. They used the age
indicator of Schweizer \& Seitzer (1992), based on morphological fine
structure. Disturbed morphology is not an accurate indicator of age, since
it depends on the details of the merger event that the progenitor galaxies
went through. OFP01a investigated normalised X-ray luminosity
(Log(L$_X$/L$_B$) versus spectroscopically determined age drawn from the
Age Catalogue of TF02, for a sample of 42 early-type galaxies. They
confirmed the trend with age. This trend has large scatter, the cause of
which is not clearly understood. There are many ideas about why there
should be such a large scatter in Log(L$_X$/L$_B$), including tidal
interactions, ram-pressure stripping, different states of ISM dynamics
(inflows, outflows or winds) that would strongly affect the X-ray
luminosity (e.g. D'Ercole, Recchi \& Ciotti 2000).  Here this trend is
revisited, also incorporating data for ongoing mergers plus some data from
more recent X-ray missions, including the three galaxies studied in this
paper. By including pre- and post-merger systems in this analysis, the
evolution of the X-ray emission can be better set into context with the
behaviour of the progenitor galaxies.

\subsection{Data and sources}
The compilations of data from OFP01b and TF02 are used, giving data for 83
early-type galaxies with known X-ray luminosities (or upper limits) and 
spectroscopic age estimates respectively. The X-ray luminosity is a 
bolometric luminosity, extrapolated from fits mainly to ROSAT data and 
assuming a 1keV MEKAL model with solar abundance. Thus these X-ray 
luminosities assume the galaxies to be dominated by a thermal component. 
They are corrected for Galactic absorption. The spectroscopic age estimate is 
a luminosity weighted average age of the stars in a galaxy. Spectroscopic age
indicators have the advantage of being able to probe back somewhat further
in time than dynamical or morphological age indicators ($\sim$several Gyr
as opposed to $\sim$ 1 Gyr), however, they have the drawback of only being
sensitive to the age of the stars, therefore a recent merger of stellar
systems in which no new star formation took place would not be detected
this way. Ages estimated from spectroscopic absorption line strengths are 
difficult to use on system younger than about 1Gyr, since the contamination
from warm gas emission is often too great.  TF02 systematically estimated
ages for a large number of galaxies, using only four spectral
line-strengths. Proctor \& Sansom (2002) showed that accuracy of age
estimates could be improved by using many spectral features at once.
However, there are few such observations.  Therefore we include here the
large number of galaxies from the above compilations.

In addition to E and S0 galaxies, data for ongoing mergers with measured 
X-ray luminosities and estimated ages were compiled (Table~7). The ages 
of these galaxies were estimated from dynamical and morphological indicators.
These age indicators are useful for young systems, but they fade over 
timescales of $<2$ Gyrs (e.g. Sansom, Reid \& Boisson 1988; Schweizer 1998; 
Brown et al. 2000). Read \& Ponman (1998, hereafter RP98) measure X-ray 
luminosities in 8 ongoing mergers. They give 
age estimates for these systems relative to zero at the time of nuclear 
coalescence, characterised by the galaxy Arp220. Therefore galaxies at 
earlier stages of merging than this have negative ages in Table~7.
They measure X-ray properties from ROSAT observations.

Fricke \& Papaderos (1999, hereafter FP99) discuss the X-ray emission from
22 interacting systems based on ROSAT data, including some systems from
RP98.  They set them onto a qualitative merging sequence.  In Table~7 we
have assigned approximate ages based on this sequence and its relation to
the age estimates of RP98. These age estimates are only approximate, but
span about the right range in pre- and post-merger ages, when guided by the
systems such as Arp270, NGC 4038/9, Arp220 and NGC7252 which have been
modelled dynamically (Mihos, Bothun \& Richstone 1993; Hibbard \& Mihos
1995; Mihos \& Hernquist 1996; Mihos, Dubinsky \& Hernquist 1998).  A colon
indicates these age estimates from ordering in FP99, rather than absolute 
age estimates. As a check on ages, in Table~7 we also indicate any cases 
with dynamical age estimates from Xilouris et al. (2004a). We find that the
estimates generally agree well, typically within 0.1 Gyr. NGC 7252 is
estimated to be somewhat younger by Xilouris et al. than in RP98, but given
the inherent errors involved, the difference is probably not significant.

We have renormalised the Log(L$_X$/L$_B$) values from FP99
by adding a constant to their plotted values. This is to try to match our 
assumption about the blue luminosity of the Sun, as used for all the other 
data points plotted in Fig.~7. This correction is explained next. It is
uncertain since it is not clear what values were used initially by the 
authors whose results we are using. In table 1 of RP98, they give large 
values for Log(L$_B$) in units of ergs s$^{-1}$, for which they appear to 
have assumed that the blue luminosity of the Sun is the same as the total 
luminosity of the Sun. This incorrect assumption appears to partly translate 
to the Log(L$_X$/L$_B$) values given in FP99. We have attempted to correct 
for this, to get all 
Log(L$_X$/L$_B$) values onto a correct dimensionless scale. To illustrate 
this correction, we take $Log(L_B(L_{B\odot}))=12.192-0.4B_T+2Log(D)$ 
for $D$ in Mpc and $M_{B\odot}=+5.48$. Then, assuming that the B band 
luminosity of the Sun is $L_{B\odot}=5.2\times 10^{32}$ erg s$^{-1}$ 
(OFP01b), we can extract optical luminosities in erg s$^{-1}$ for 
the galaxies. Doing this, and comparing with the B band
luminosities and Log(L$_X$/L$_B$) ratios given in RP98, and 
FP99, we estimate that a correction of +0.61 is required 
to the Log(L$_X$/L$_B$) values plotted in FP99, to get them 
onto a correct scale. In Table~7 we have applied this correction.

Keel \& Wu (1995) estimated dynamical ages for 35 ongoing merger candidates.
They put them into an evolutionary sequence and estimated dynamical stages 
in terms of crossing times, using morphology and kinematics. A literature 
search reveals that their sample is not well studied in X-rays, therefore 
we cannot incorporate any more cases from Keel \& Wu into this present 
study. However for 8 systems in common their ordering is the same as in 
Table~7, providing independent support for the age ordering given in Table~7.

\bigskip
\begin{table*}
\centering
\begin{minipage}{150mm}
\caption{\bf Data for ongoing merging and interacting galaxies with 
dynamical age estimates and X-ray measurements from Read \& Ponman 1998 
(RP98) and Fricke \& Papaderos 1999 (FP99). The X-rays were measured in 
the range 0.1 - 2.4 keV, from ROSAT observations. Comparison ages are 
given for 6 overlapping cases from Xilouris et al. 2004a (Xea04).
See Section 5.1 for discussion of data sources.}
 
\begin{tabular}{lllc}
 & & & \\
{\bf Galaxy} &{\bf Age} &{\bf Ref.} &{\bf $Log({{L_X}\over{L_B}})$} \\
 & {\bf (Gyr)} & &  \\
 & & & \\
NGC 2342  &-0.85:   &FP99 &-2.29 \\
NGC 2341  &-0.8:    &FP99 &-2.09 \\
NGC 2993  &-0.75:   &FP99 &-1.86 \\
Arp 102a  &-0.7:    &FP99 &-1.89 \\
Arp 284   &-0.65:   &FP99 &-1.84 \\
Arp 270   &-0.6     &RP98 &-2.69 \\
Arp 242   &-0.5     &RP98 &-2.34 \\
Arp 299   &-0.47:   &FP99 &-2.29 \\
Mk 1027   &-0.43:   &FP99 &-2.09 \\
NGC 4038/9  &-0.4/-0.25   &RP98/Xea04 &-2.19 \\
Arp 278   &-0.3:    &FP99 &-2.41 \\
NGC 520   &-0.2/-0.19     &RP98/Xea04 &-2.99 \\
Arp 215   &-0.171:  &FP99 &-2.29 \\
NGC 3310  &-0.143:  &FP99 &-2.19 \\
Mk 789   &-0.114:   &FP99 &-2.09 \\
Mk 266   &-0.086:   &FP99 &-1.69 \\
NGC 6240  &-0.057:/-0.03  &FP99/Xea04 &-1.29 \\
Mk 231   &-0.029:   &FP99 &-1.84 \\
Arp 220   &0.0/0.0        &RP98/Xea04 &-2.19 \\
NGC 2623   &0.1/0.16      &RP98/Xea04 &-2.24 \\
NGC 7252   &1.0/0.24      &RP98/Xea04 &-2.94 \\
AM1146-270   &1.5   &RP98 &-2.69 \\
\multicolumn{4}{l}{A colon indicates approximate ages from ordering in FP99} \\
\end{tabular}\\
\end{minipage}
\end{table*}

\bigskip
\begin{table*}
\centering
\begin{minipage}{150mm}
\caption{\bf Compiled data for galaxies with accurate published age estimates 
and/or recent X-ray measurements from the literature. Where possible,
X-ray fluxes are ones corrected for absorption in our Galaxy. Distances 
(D) are from Prugniel \& Simien (1996). Apparent total B magnitudes (B$_T$) 
are from RC3. Other sources of information are indicated.See Section 5.1 
for a discussion of data sources.}
\begin{tabular}{lcccccccc}
 & & & & & & & & \\
{\bf Galaxy} &{\bf D} &{\bf Age} &{\bf Ref.} &{\bf B$_T$} &{\bf X-ray flux} 
&{\bf $Log({{L_X}\over{L_B}})$} &{\bf X-ray band} &{\bf Ref.} \\
 &{\bf (Mpc)} &{\bf (Gyr)} &{\bf (Ages)} &{\bf (mag.)} 
&{\bf (erg s$^{-1}$ cm$^{-2}$)} & &{\bf (keV)} &{\bf (X-rays)} \\
 & & & & & & & & \\
NGC 4382  &15.9       &1.6       &TF02        &10.00     &$8.66\times 10^{-13}$    & -2.89 &0.3-7.0     &This work \\
          &           &          &            &          &$10.75\times 10^{-13}$   & -2.80 &0.3-10.0    &SSI03 \\
NGC 5363  &15.8$^a$   &6.7       &This work   &11.05     &$4.78\times 10^{-13}$    & -2.73 &0.3-7.0     &This work \\
NGC 2865  &36.5    &1.0/$<$1.5   &Hea99/TF02  &12.57     &$1.00\times 10^{-13}$    & -2.80 &0.5-10.0     &This work \\
NGC 4365  &15.9       &9.7       &PS02        &10.52     &$8.87\times 10^{-13}$    & -2.67 &0.3-10.0    &SSI03 \\
NGC 3585  &16.07$^c$  &3.1       &TF02        &10.88     &$1.44\times 10^{-13}$    & -3.32 &0.2-8.0     &OP04a \\
NGC 4494  &21.28$^c$  &7.5/6.7   &This/Dea05  &10.71     &$2.34\times 10^{-13}$    & -3.18 &0.2-8.0     &OP04a \\
NGC 5322  &27.80      &4.2/2.4   &PS02/Dea05  &11.14     &$2.87\times 10^{-13}$    & -2.92 &0.2-8.0     &OP04a \\
NGC 3921  &72.8       &0.7       &S96         &13.06     &$3.68\times 10^{-13}$    & -2.04 &0.5-10.0    &Nea04 \\
NGC 7252  &52.5       &1.0/0.8   &RP98/HM95   &12.72     &$1.28\times 10^{-13}$    & -2.64 &0.5-10.0    &Nea04 \\
NGC 1600  &59.98      &6.9       &TF02        &11.93     &$1.23\times 10^{-12}$    & -1.97 &0.3-10.0    &SSC04 \\
NGC 1700  &50.58      &2.3       &TF02        &12.20     &$2.64\times 10^{-13}$    & -2.53 &0.3-2.7     &SM02 \\
NGC 4636  &15.9       &8.2       &PS02        &10.43     &$2.81\times 10^{-11}$    & -1.21 &0.5-4.0     &Mea98 \\ 
NGC 5102  &3.1        &3.0       &Kea05       &10.35     &$5.17\times 10^{-14}$    & -3.98 &Bolometric  &OFP01b \\
NGC 4473  &16.14      &9.4       &CRC03       &11.16     &$5.77\times 10^{-13}$    & -2.61 &Bolometric  &OFP01b \\
NGC 4621  &15.92      &17.3      &CRC03       &10.57     &$3.45\times 10^{-13}$    & -3.06 &Bolometric  &OFP01b \\
NGC 3256  &35.4      &$\sim 0.0$ &Jea04       &12.15     &$1.00\times 10^{-12}$    & -1.97 &0.3-10.0    &Jea04 \\     
 & & & & & & & \\
\end{tabular}\\
TF02 = Terlevich \& Forbes (2002) \\
Hea99 = Hau et al. (1999) \\
PS02 = Proctor \& Sansom (2002) \\
Dea05 = Denicol\'{o} et al. (2005) \\
S96 = Schweizer (1996) \\
HM95 = Hibbard \& Mihos (1995) \\
Kea05 = Kraft et al. (2005) (for age and distance) \\
CRC03 = Caldwell, Rose \& Concannon (2003) \\
Jea04 = Jenkins et al. (2004) \\
SSI03 = Sivakoff, Sarazin \& Irwin (2003) \\
OP04a = O'Sullivan \& Ponman (2004a) \\
Nea04 = Nolan et al. (2004) (nuclear regions plus hot diffuse gas) \\
SSC04 = Sivakoff, Sarazin \& Carlin (2004) (unresolved sources plus gas) \\
SM02 = Statler \& McNamara (2002) \\
Mea98 = Matsushita et al. (1998) \\
$^a$ = Distance from Tully (1988) \\
$^c$ = Distances incorrectly reversed in OP04a \\
\end{minipage}
\end{table*}

The data for early-type galaxies from OFP01b is supplemented here with more
recent X-ray data for some galaxies, and for some galaxies with more 
accurately measured ages.
These results from more recent observations for some systems are 
indicated in Table~8. No attempt has been made to correct for the different 
X-ray wavebands since such corrections are very dependent on the accuracy of 
fitted parameters, especially when extrapolating to broader wavebands. 
Instead we have tried to include results quoted from 
broad wavebands, where possible. 

Of the three early-type galaxies analysed in this paper, NGC 4382 is well 
established as having a young, luminosity weighted age 
(1.6 Gyr $\pm$ 0.3) from a fit to four optical spectral line-strengths 
(H$\beta$ and the combination index [MgFe]) by TF02. 
This galaxy has strong H$\beta$ and higher order Balmer lines in 
absorption, indicative of a young stellar population. 
Similarly NGC 2865 has strong H$\beta$ and a young age (Hau et al. 1999).
NGC 5363 was originally thought to be another very young early-type galaxy, 
from the strength of its H$\beta$ absorption. Denicol\'{o} et al. (2005) 
estimated its age from fits utilising 4 indices and found 
3.8$^{+2.1}_{-3.5}$ Gyr. To try to reduce the uncertainty 
on this age estimate (and age estimates for other galaxies in their sample) 
we fitted many more optical line-strengths, from the data of Denicol\'{o} 
et al. For NGC 5363 we found 6.7 $\pm$ 0.5 
Gyrs. Therefore this galaxy, although relatively young, is not as young as 
originally thought. This gives a typical illustration of the inherent 
difficulties in determining spectroscopic ages of galaxies (e.g. 
Trager 2004). In the current compilation of data (Table~8, Fig.~7 and Fig.~8) 
we use this latter age estimate. Total unabsorbed X-ray fluxes are given 
in Table~8.

There are 23 other galaxies in the Denicol\'{o} et al. sample that have X-ray 
measurements. We fitted these, using the line-strength indices measured 
by Denicol\'{o} et al. to determine accurate luminosity weighted ages.
Between $\sim 10$ to 17 optical indices were fitted with single-age, 
single-metallicity stellar population models of Thomas, Maraston \& Korn 
2004. The $\chi^2$ statistic was minimised to derive luminosity weighted 
ages, metallicities and $\alpha$-element abundance ratios. The rational 
for using as many spectral line-strengths as possible is that, while all 
indices show some degeneracy with respect to age and metallicity, each 
index contains some information regarding each parameter.

For fitting the Denicol\'{o} et al. sample, indices redward of 
Fe5046 were generally excluded. These indices are often problematic for a 
variety of observational reasons (e.g. inter-stellar absorption in NaD, 
flux calibration issues for TiO indices) and, in this case, showed large 
residuals to the best fits. The poorly modelled G4300 was also excluded 
for similar reasons. For the remaining indices an (approximately) 3-sigma 
clipping process was employed. 
This resulted in, typically, 1 or 2 indices per galaxy being removed 
from the fitting procedure (an average of 1.6 indices per galaxy).
Of these, more than 50\% were associated with known problems (e.g.
emission-line filling of the H$\beta$ index, low sensitivity and poor
sky-subtraction in indices redward of $\sim$5500~\AA\ and the
flux-calibration sensitivity of the Mg$_1$ and Mg$_2$ indices). In any
case, the derived log(age) and metallicities were highly robust with
respect to the clipping procedure, generally changing by no more than
$\sim$0.1 dex from the values obtained when all available indices are
included (i.e. with no clipping). See Proctor et al. 2004 for details of
the clipping procedure. The resultant fits therefore typically 
included between 10 and 17 indices (note Denicol\'{o} did not measure all 
indices, in all galaxies) and showed reduced-$\chi^2$ values of order 1.5. 
Errors were derived from 100 Monte Carlo realisations of the best fit model 
data. These newly derived, luminosity weighted ages are plotted 
in Figs.~7 and 8.

Other early-type galaxies have been observed with recent X-ray missions.
Sivakoff, Sarazin \& Irwin (2003) observed two X-ray faint ellipticals 
(NGC 4382 and NGC 4365) with Chandra. Total fluxes were obtained from 
their Tables 3 and 4, within 6r$_e$ diameter, assuming a power-law for the 
hard component. Published results also include O'Sullivan \& Ponman (2004a), 
who analysed XMM-Newton and Chandra data for three X-ray faint, 
early-type galaxies (NGC 3585, 4494, 5322). Two merger-remnant galaxies, 
NGC 3921 and NGC 7252, were observed by Nolan et al. (2004), with XMM-Newton.
They give luminosities for nuclear regions, extended hot gas, and other 
X-ray point sources. In Table~8 we have combined their nuclear and 
extended hot gas components to determine an overall flux from each of 
these two galaxies. Sivakoff, Sarazin \& Carlin (2004) observed NGC 1600
with Chandra. We have summed their hot gas plus unresolved source components
to get the overall X-ray flux from this galaxy, as given in Table~8. 
Statler \& McNamara (2002) obtained Chandra observations of the extended 
disk-like, X-ray structure in the elliptical galaxy NGC 1700. The total 
flux from their work is given in Table~8. Deep exposure observation of 
NGC 4636 were taken with the ASCA satellite by Matsushita et al. (1998). 
NGC 4636 is an X-ray luminous elliptical galaxy with very extended X-ray 
emission. They fit two $\beta$ components to the Chandra surface brightness
profile for NGC 4636, out to a radius of 60 arcminutes and give the 
total luminosity in the broader $\beta$ component as 
8.1$\times 10^{41}$ erg s$^{-1}$. They say that this exceeds their 
compact $\beta$ component by a factor of 5. In Table~8 we have summed 
the flux from these two beta components, for NGC 4636. 
New estimates of spectroscopic ages are available for the lenticular galaxy 
NGC 5102 (Kraft et al. 2005) and two galaxies from the sample of Caldwell, 
Rose \& Concannon (2003) (NGC 4473 and NGC 4621). Jenkins et al. (2004) 
obtained XMM-Newton observations of the starburst merger galaxy NGC 3256, 
which is thought to be the product of two gas-rich galaxies of roughly 
equal size, from the morphology and ongoing star formation. It is at a 
similar evolutionary stage to that of Arp 220. The X-ray flux 
and age estimate for NGC 3256 is included in Table~8, from Jenkins et al.
The data in Table~8 are included in Fig.~7 and Fig.~8.

\begin{figure*}
\begin{minipage}{150mm}
\includegraphics[angle=-90,width=165mm,trim=10 0 0 0]{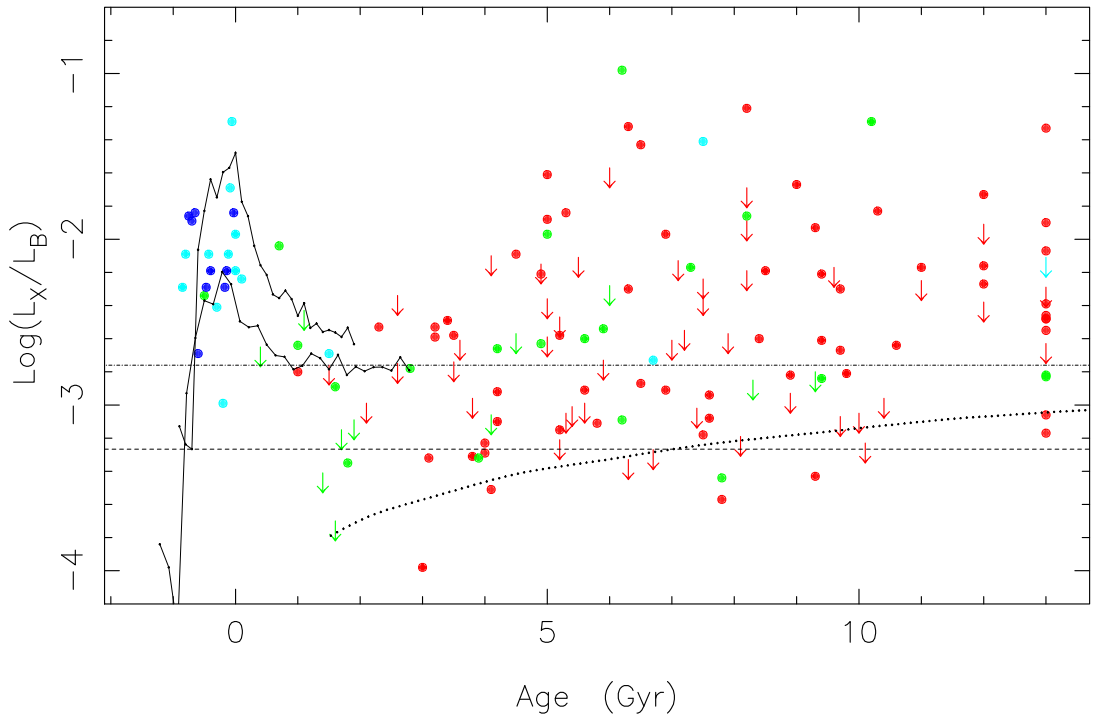}
 \caption{Normalised X-ray luminosity versus age for ongoing mergers and 
early-type galaxies, including data from Table~7 and 8. Age estimates greater 
than 13 Gyrs are plotted at 13 Gyrs to avoid unrealistically stretching 
the horizontal scale in this plot. $3\sigma$ upper limits to the X-ray 
emission are indicated by downward arrows. The dashed and dot-dashed 
straight lines represent the ratio expected from stellar sources only 
(LMXBs) from OFP01b and Kim \& Fabbiano (2004) respectively.
The dotted curve illustrates the effect on the stellar contributions due to 
optical fading of a starburst at age zero. This curve is arbitrarily 
normalised to the lower straight line at age 7 Gyr.
Also shown are two curves (solid lines) representing merger models of T.Cox 
(see text for details). The blue band luminosity of the Sun was taken to be 
5.2$\times 10^{32}$ erg s$^{-1}$, from OFP01a. The colour coding indicates 
Hubble type: Elliptical (red), S0 (light green), Spiral (dark blue), Irregular 
(turquoise).}
\end{minipage}
\end{figure*}

Predictions of X-ray luminosity changes through the evolution of a major 
merger of two spiral galaxies have been made by Cox et al. (2005). We use 
their simulation of X-ray luminosity over 3 Gyrs (data provided by T.Cox, 
private communication) 
to estimate total Log(L$_X$/L$_B$) versus age. Zero age corresponds to 
1.1 Gyrs in their simulation, since this is the time of nuclear coalescence. 
We have taken L$_B$ to be constant with age and equal to the average value 
for the galaxies plotted in Fig.~7 at $<2$ Gyr (L$_B$=8.8309$\times 10^{42}$ 
erg s$^{-1}$). A constant discrete source contribution is added to the X-ray 
emission, estimated at L$_{X(dscr)}=4.78\times 10^{39}$ erg s$^{-1}$, from 
Log(L$_{X(dscr)}$/L$_B$)=29.45 (erg s$^{-1}$ L$_{B\odot}$) given in OFP01b. 
This discrete source contribution was estimated from the lower envelope of 
L$_X$ versus L$_B$ emission in early-type galaxies, assuming unity slope 
(see OFP01b for details).
Although the optical luminosity is unlikely to remain constant over the 
3 Gyrs pre- and post-merger, we detected no strong systematic change in 
L$_B$ with age, for our plotted galaxies, as there is a large spread in 
L$_B$ at all ages in this sample. The discrete X-ray source 
contribution is also likely to vary somewhat over this time, but we 
currently have little information about this. We know that high-mass 
X-ray binaries (HMXBs) should give a boost to L$_X$ around the time 
of the merger, but the evolution of the LMXB population is not known.
A more systematic spectral analysis of the X-ray data would be required 
to investigate this issue, incorporating two components for many galaxies.
The colour coding in Fig.~7 indicates the different Hubble types. A dearth 
of luminous X-ray sources is evident for both elliptical and S0 
classifications, in the post-merger age range of 1 to 4 Gyrs.

The model curve plotted 
in Fig.~7 (upper solid line) fits reasonably well to the ongoing merger data, 
and predicts slightly more hot gas than we detect in the data at post-merger 
ages. Another of the simulations from Cox et al. is also shown in Fig.~7, 
covering 4.3 Gyrs, with a peak in X-rays at about 1.5 Gyrs into the simulation. 
In this example we were also able to account for changes in the blue 
luminosity with time (data provided by T. Cox, private communication). 
This is the lower of the two model curves (solid lines)
plotted in Fig.~7. Accounting for temporal changes in the blue luminosity 
reduces the peak below the observed data and predicts a very low 
Log(L$_X$/L$_B$) in the earliest stages of the merger. Cox et al. 2005 find 
low (L$_X$/L$_B$) for their model post-merger when compared to elliptical 
galaxies of similar L$_B$. In either example 
model the post-merger values are comparable with the highest observed 
detections around those times ($<3$ Gyrs after coalescence). Therefore 
the models appear to slightly over predict the hot gas components present in 
post-merger galaxies.

Thus the observations can constrain the reality of models. The range 
of possible model behaviours needs further investigation, plus extension 
to older systems, to test the expected X-ray evolution of galaxies resulting 
from mergers. 

\begin{figure*}
\begin{minipage}{150mm}
\includegraphics[angle=-90,width=165mm,trim=10 0 0 0]{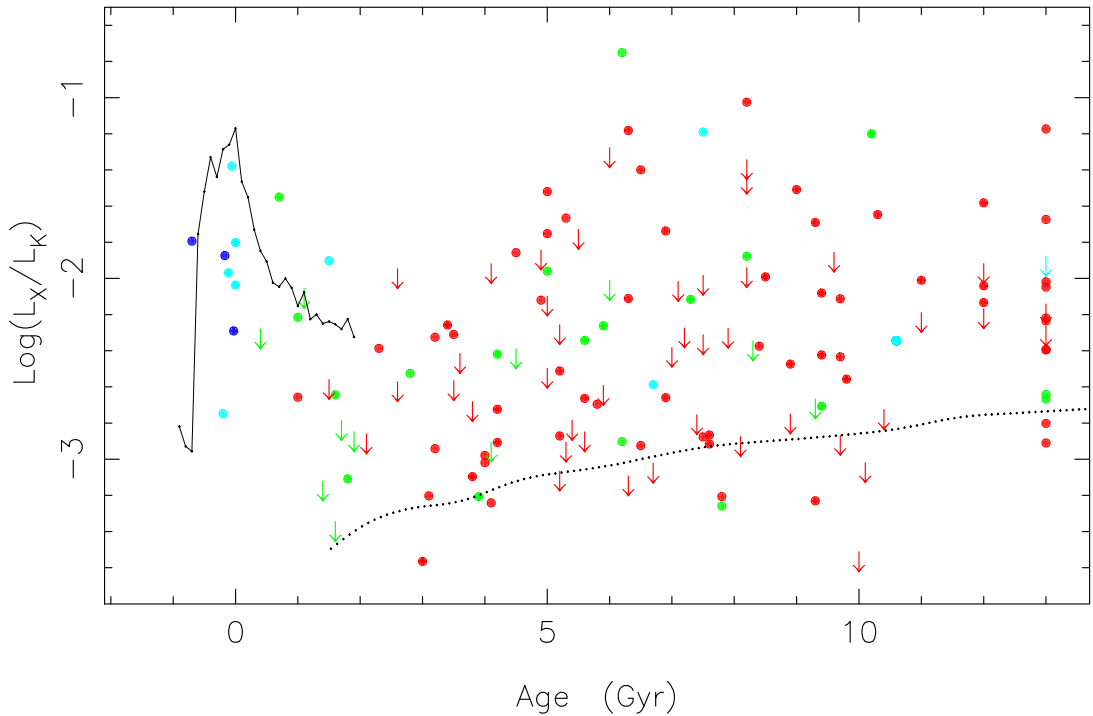}
 \caption{Normalised X-ray luminosity versus age including data from Table~8,
as in Fig.~7, except normalised by the K band luminosity, from 2MASS data, 
where available. The colour coding is the same as in Fig.~7. X-ray evolution 
from the model of T.Cox is shown, assuming a constant K-band luminosity.
The dotted curve illustrates the effect on the stellar contributions due to 
K-band fading of a starburst at age zero. This curve is arbitrarily 
normalised to the level expected from discrete X-ray sources at age 7 Gyr,
estimated using average B and K band luminosities (Log(L$_B$)=42.90, 
Log(L$_K$)=42.60 erg s$^{-1}$) from the galaxies plotted.
}
\end{minipage}
\end{figure*}

The B band luminosity is very sensitive to young stars. The K band
is more representative of the underlying stellar mass in a galaxy.
Therefore in Fig.~8 we have plotted the X-ray luminosity normalised by the 
K band luminosity. We used total K band magnitudes from the 2MASS survey, 
where available. For the 2MASS K band we assume (V-K)$_{\odot}$=1.45
(Toft, Soucail \& Hjorth 2003), M$_{V\odot}$=+4.82, and a zero point 
flux calibration of K=0 for a flux of 
F$_K$=1.122$\times 10^{-14}$ J s$^{-1}$ cm$^{-2}$ (Cohen et al. 2003). 
This gives an apparent K-band magnitude of K$_{\odot}$=-28.20 and a 
luminosity of L$_{K\odot}$=6.03$\times 10^{31}$ erg s$^{-1}$.
Predicted X-ray evolution is shown (Cox et al. 2005), assuming a 
constant L$_K=4.333\times 10^{42}$ erg s$^{-1}$ estimated from the 
galaxy data at $<2$ Gyr. This prediction also includes an estimate 
of the discrete source contribution to the X-rays.
Fig.~8 shows a similar trend as seen in Fig.~7, confirming
the relatively low X-ray flux levels in post-merger systems.
Note that Figs.~7 and 8 exclude the exceptionally X-ray bright cD galaxy,
IC 5358, which has a spectroscopic age of 16 Gyr and Log(L$_X$/L$_B$)=0.0.

\subsection{Interpretation of trends with age}
The horizontal lines in Fig.~7 give an indication of the level expected 
from stellar
contributions only and its uncertainty. Within this uncertainty Fig.~7
indicates very little hot gas in E and S0 galaxies younger than about 4
Gyrs. Massive hot gas halos only appear to occur in older early-type
galaxies. In Figs.~7 and 8 we have plotted a fading starburst, using the 
code described in Sansom \& Proctor (1998).
The level of optical fading even from a short, intense burst, is
insufficient to account for this result. The evolution of the X-ray
luminosity of the X-ray binary population is not well understood. HMXBs
will dominate in the few hundred Myr after the merger, and
probably help to produce the peak in X-ray luminosity seen around Age=0,
but their contribution will rapidly decline with the population of high
mass stars. In the long term, LMXBs will
dominate, but the evolution of the integrated luminosity of the population
over many Gyr is not well understood. A factor $\sim$6 increase would be
required to explain the general trend we see, and this seems unlikely.
Note that ages greater than about 10 Gyrs are very uncertain and some 
may be biased by emission line filling. Any galaxies with estimated ages 
greater than 13 Gyrs are therefore plotted at 13 Gyrs in Fig.~7 
and Fig.~8 to avoid stretching the age scale unrealistically. 
Most of the galaxies in OFP01b had ages uncertain by $\sim 20$\%.

The presence of an accreting central black hole may provide a good explanation 
of the gas-poor state of post-merger galaxies, since it can act to drive a 
strong wind, as shown in the model of Cox et la. 2005. If the wind is 
sufficient to expel the gas then the halo building stage will be delayed 
until enough new ISM gas is built up from stellar mass loss. 
Return of gas from post-merger tidal features (modelled by Hibbard 
\& Mihos 1995) cannot account for the hot gas halos since post-merger 
galaxies are both X-ray poor (see Figs.~7 and 8) and show little evidence 
for cold gas (Sansom et al. 2000). Therefore hot gas halos may instead 
be built up from stellar mass loss in the post-merger phase. This may 
lead to a more metal enriched ISM than in the case of returning, 
pre-existing gas, since the mass loss would be from more metal enriched stars.

Future requirements to test this picture will include examining hot gas 
properties and stellar X-ray components separately, versus age. This can 
only be done when accurate and consistent X-ray spectral fits have been 
made for enough cases, from data 
covering a broad waveband to avoid uncertain extrapolations. Data mining
XMM-Newton and Chandra archives will allow this to be done. To date, most 
published papers on large galaxy samples, observed using these satellites, 
do not give sufficient information to do this (e.g. White et al. 2002; 
Diehl \& Statler 2005). Some such data are starting to appear 
(e.g. Fukazawa et al. 2006), but more is needed to map properties with age.
The effects of environment have not been considered here, but this should
also be important to do for a more homogeneous data set in future.

\section{Conclusions}

Three early-type galaxies, with young stellar populations, were observed with 
XMM-Newton or Chandra. Two (NGC 5363 and NGC 2865) are dominated by stellar 
contributions to the X-ray emission and the other (NGC 4382) has roughly 
equal flux contributions from stellar and hot gas components. Thus we do 
detect low levels of hot gas in these spectroscopically young, early-type 
galaxies, confirming previously published results. A revised, older age was
found from re-analysis of optical spectra for NGC 5363, using more spectral 
indices.

An attempt was made to recover the mass distribution in NGC 4382, which
indicates an extended mass distribution, not bounded by the optical or
X-ray light, however the gaseous halo is only detected out to $\sim 16$ kpc
($\sim$ 4r$_e$) radius. In NGC 4382 there is a gas mass of 
$\sim 4\times 10^8$ M$_{\odot}$ within this radius. In contrast,
NGC 5363 is at a similar distance, but has about a factor of three less 
flux in the thermal component, so is likely to possess correspondingly 
less hot gas. Therefore there is no strong evidence for the quantities 
of gas expected from the observed dust mass, assuming gas-to-dust mass 
ratios typically measured for early-type galaxies.
For NGC 4382 the mass-to-light ratio in the galaxy core is consistent with 
stars being the dominant form of mass, but the mass profile rises with radius,
suggesting an increasing dark matter contribution. Therefore this 
spectroscopically youthful system appears to possess a dark matter halo.

Data from the literature, together with new data presented in this paper, 
were compiled for merging and early-type galaxies with measured, normalised 
X-ray emission (Log(L$_X$/L$_B$)) and estimated ages. This compilation of data 
confirms that there is a drop in X-ray emission for relatively young, 
early-type galaxies and illustrates that this drop extends up to  
$\sim 4$ Gyr in age. This is confirmed in the Log(L$_X$/L$_K$) versus age 
plane. The most likely explanation is that wind activity, fuelled by star 
formation and possible AGN activity, causes a lack of hot gas in post 
merger galaxies. 
Future work will investigate the origin of this behaviour, through 
separate X-ray spectral components, for a large, uniformly analysed sample.

\section*{Acknowledgments}

We thank A.Read for advice on XMM-Newton analysis and B. Maughan for 
use of his sigma-clipping software. We thank T. Ponman for help obtaining 
the NGC 2865 data and T. Cox for sending us results of his simulations prior
to publication. AES would like to thank the University
of Central Lancashire for awarding a research sabbatical to complete this 
work. EOS acknowledges support from NASA grant NNG04GM97G.
The authors made use of the NASA Extragalactic Data base. 
Thanks go to an anonymous referee for significant improvements in this paper.

\label{lastpage}

\end{document}